\documentclass[useAMS,usenatbib]{mn2e}

\usepackage[pdftex]{graphicx}

\newcommand{\ngca}{$N_{GC}$}
\newcommand{\mbha}{$M_{\bullet}$}
\newcommand{\rea}{$R _{e}$}
\newcommand{\sigea}{$\sigma _{e}$}
\newcommand{\mdyna}{$M_{dyn}$}

\newcommand{\sige}{$\sigma _{e}~$}
\newcommand{\re}{$R _{e}~$}
\newcommand{\ngc}{$N_{GC}~$}
\newcommand{\mbh}{$M_{\bullet}~$}
\newcommand{\mdyn}{$M_{dyn}~$}
\newcommand{\mue}{$\mu_{e}~$}

\def\lta{\mathrel{\spose{\lower 3pt\hbox{$\mathchar"218$}}
     \raise 2.0pt\hbox{$\mathchar"13C$}}}
\def\gta{\mathrel{\spose{\lower 3pt\hbox{$\mathchar"218$}}
     \raise 2.0pt\hbox{$\mathchar"13E$}}}

\title[GCs and SMBHs in galaxies: a larger sample]
{Globular clusters and supermassive black holes in galaxies: 
further analysis and a larger sample}
\author[Gretchen L. H. Harris, Gregory B. Poole, and William E. Harris] 
{Gretchen L. H. Harris$^{1}$\thanks{E-mail: glharris@astro.uwaterloo.ca;
 gregory.poole@unimelb.edu.au; harris@physics.mcmaster.ca}
Gregory B. Poole$^{2}$ 
and William E. Harris$^{3}$ \\ 
$^{1}$Department of Physics and Astronomy, University of Waterloo, 
Waterloo ON N2L 3G1, Canada \\
$^{2}$School of Physics, University of Melbourne, Parkville, Victoria 3010,  Australia \\
$^{3}$Department of Physics and Astronomy, McMaster University, Hamilton ON L8S 4M1, Canada}
\begin{document}
\date{Accepted  Received; in original form }

\pagerange{\pageref{firstpage}--\pageref{lastpage}} \pubyear{2002}

\maketitle

\label{firstpage}

\begin{abstract}
We explore several correlations between various large-scale galaxy properties, 
particularly total globular cluster population (\ngca), the central black hole mass
(\mbha), velocity dispersion (nominally \sigea), and bulge mass (\mdyna). 
Our data sample of 49 galaxies, for which both \ngc and \mbh are
known, is larger than used in previous discussions 
of these two parameters
and we employ the same sample to explore all pairs of correlations.
Further, within this galaxy sample we investigate the scatter in
each quantity, with emphasis on the range of
published values for \sige and effective radius (\rea) for any one galaxy. 
We find that these two quantities in particular are 
difficult to measure consistently and caution that precise intercomparison
of galaxy properties involving \rea and \sigea is particularly difficult. 


Using both conventional $\chi^2$ minimization and Monte Carlo Markov Chain (MCMC) fitting techniques, 
we show that quoted observational uncertainties for all parameters are too 
small to represent the true scatter in the data. 
We find that the correlation between \mdyn and \ngc is stronger than 
either the \mbh - \sige or the \mbh - \ngc relations.  We suggest that 
this is because both the galaxy bulge population
 and \ngc were fundamentally established at an early epoch
during the same series of star-forming events.
By contrast, although the seed for \mbh was likely formed at a similar epoch, its growth over
time is less similar from galaxy to galaxy and thus less predictable.

\end{abstract}


\section{Introduction}
 
As the available database describing the global properties of individual galaxies has  
grown, it has become increasingly feasible to try to understand more about these
systems by exploring correlations among several large-scale 
structural parameters.  These global parameters often include: 
\sige (bulge velocity dispersion), 
\mdyn (bulge mass as calculated through the virial theorem or its variants), \re (effective radius), 
\mue (mean surface brightness), luminosity, and more recently
\mbh (mass of the central supermassive black hole). 
One of the earliest demonstrations of such correlations was the work of \citet{d2} 
 who used a sample  of 260 systems to show that early-type galaxies form a two-parameter 
family characterized by velocity dispersion and mean surface brightness, introducing the 
concept of a fundamental plane for galaxies.  Correlations with \mbh 
properties were first constructed by \citet {k3}  
 (8 galaxies) and \citet{m5} (36 galaxies).  
Since then, \citet{f4} and \citet{g13}
demonstrated a correlation between \mbh and \sigea, which appeared to have lower dispersion
than the relation between \mbh and the luminosity of the galactic bulge or other 
parameters.  However, the 
slopes of the \mbha - \sige relation derived by these two groups were significantly
different and the debate on this point continues energetically today 
\citep[e.g.][to cite a few]{mf01,t1,novak06,gra11,gu11}.

Recently, \citet{spit09} proposed links among a galaxy's globular cluster system (GCS),
total halo mass, and SMBH mass.   
From a sample of 13 galaxies, \citet {b1} (hereafter BT) 
also showed that \mbh is almost directly proportional to the total globular cluster 
population of a galaxy, \ngca.  At first glance the statement that \ngc $\sim$ 
\mbh should not be thought of as terribly surprising, since 
both \ngc and \mbh are known to be related to other galaxy properties such as mass and 
luminosity (see especially \citet{hha13} (hereafter HHA13) for a recent comprehensive discussion
of GC populations and systematics);
essentially, larger galaxies should on average have more massive SMBHs and more
globular clusters.  But more interesting, and possibly more important for understanding
early galaxy evolution, is the suggestion that the {\sl scatter} in the \mbh - \ngc correlation is 
remarkably small, comparable with or even lower than what has been claimed for \mbh vs.~\sigea.  
The low scatter is harder to explain by trivial scaling arguments, and is
what makes this newly-found correlation intriguing.

\citet {hh} 
(hereafter HH) confirmed the findings of \cite{spit09} and BT  from a much larger sample of 33 galaxies
with well determined \ngc and \mbha~ values.  They found, as did BT but with stronger
confidence, that \mbh scales nearly linearly with \ngc (\mbh $\sim$ \ngc$^{0.98 \pm 0.10}$)
and also drew attention to a number of other issues:
\begin{itemize}
\item{} Occasional genuine outliers that fall well off the central \ngc - \mbh relation cannot be
explained by measurement uncertainties in either quantity.  Most notably, the Milky Way and NGC 5128
are both strong outliers in the sense that they have too-small SMBH's for their number
of GCs, though both quantities are well determined.  
\item{} The residual scatter around the mean relation \mbh $\sim$ \ngc is very noticeably larger
than expected from the random measurement uncertainties quoted in the literature, suggesting
that the scatter is dominated by a real cosmic variance.
\item{} The spiral galaxies (with the exception of the Milky Way) follow the E-galaxy
relation quite closely.  By contrast, the S0 galaxies in the list show much larger
scatter and no well defined slope; for them, the basic E/S relation ``appears to be irrelevant''.
\end{itemize}

Additional recent discussions of the correlations between the GCS and the SMBH 
have been published by \citet{sny11}, \citet{sad12}, and \citet{rho12}.
\citet{sny11} argue essentially that the relatively low scatter in \mbh - \ngc can
be interpreted by seeing both quantities as 
linked to (and perhaps determined by) the galaxy's bulge mass.
\citet{sad12} extend the discussion to propose a correlation between \mbh and
the velocity dispersion \emph{of the globular cluster system}, $\sigma_{GCS}$; at present,
this interesting relation can be built on only 13 large galaxies with available GCS
velocity data.

\citet{rho12} selected 20 galaxies (10E, 8S0, and 2S) 
with particularly complete, homogeneous GCS data and found
\mbh $\sim$ \ngc$^{1.20 \pm 0.06}$, 
a $\sim 20$\% steeper slope than in HH.
The steeper slope seems primarily to be due to 
her use of only large galaxies in the fit (essentially Milky-Way-sized and larger),
whereas HH used galaxies over the entire existing luminosity range, including dwarfs.  HH also include two very
luminous E's, M87 and NGC 5846, that Rhode does not.  These extreme cases at the top end
and bottom end of the correlation make up only a minority of the total sample but they
exert strong leverage over the fit.  It is not out of the question that with much
more, and more precise, data in hand the \ngc - \mbh relation might appear slightly nonlinear.
Notably, \citet{rho12} also finds that 
the slope does not change significantly if the GC sample is subdivided into its
blue (metal-poor) and red (metal-richer) subpopulations, a step that can be done
only for high-quality GCS photometry in which the division between red and blue can
be clearly made.  As in HH, she also found that the S0 and disk galaxies in this selected sample showed
larger scatter than the E's (though still based on a very small sample).

In this paper, we explore the various correlations of these global parameters
for a still larger sample of galaxies,
paying the closest attention to four of the ones listed above: \mbha, \ngca, \sigea, and 
\mdyna.  We find that the relation with the smallest scatter of
all is the one between \ngc and the dynamical bulge mass, \mdyna, and suggest a possible
evolutionary link between them.  We have two additional objectives.  First, we directly
compare best-fit solutions from a conventional 
$\chi^2-$minimization procedure, with ones obtained from a MCMC (Monte Carlo Markov Chain)
technique.  Second, we discuss the residual scatters around these solutions in an attempt
to decide whether or not we have already resolved the intrinsic (cosmic) scatter in the
parameters.

\section{Causal Connections?}

The SMBH of a galaxy is a structural feature with a typical size of a few AU, whereas the 
GCS spans tens to hundreds of kiloparsecs, a factor of $10^{10}$ bigger in scale size.
So a \emph{causal} link between the SMBH and the globular cluster population 
may therefore seem quite unlikely. But the possibility should not simply be dismissed out of hand.
HH pointed out that both the SMBH and GCS have their origins in the same earliest phases of
galaxy formation at $z \sim 5 - 10$.  
There is a rich literature on how SMBH-driven AGN activity can strongly influence star formation
in large galaxies through either negative or positive feedback, 
particularly in the high-redshift stages when the most gas is present \citep{sil98,fab12}.
To cite only a few recent examples, \citet{wag12,wag13} present models of AGN jets interacting
with a fractal-like ISM in the central kpc. These show that a high fraction of the jet energy
is transferred to the gas, and also that the jet can quickly ($< 10^6$ y) penetrate through the central bulge
out into the halo regions by channeling through the intercloud regions.  Lower-density regions
of the gas are ablated and heated by the jet, but the densest gas cores (among which would be protoglobular
clusters) are preserved.  In addition, the star formation
rate outside the central region will be enhanced through AGN ram compression
\citep{fab12,gai12,sil05,ish12}.
Once the powerful jet penetrates into the halo, its energy can be deposited and
percolated over many-kiloparsec scales within a few $10^8$ years, an appropriate
time period for the formation of the first GCs.

On the observational side, AGN- or QSO-driven star formation in galaxies at
$z \sim 2-3$ seems able to occur in sites many tens of kpc from the galaxy center
\citep[e.g.][among many others]{rau13,cro06,kla04,bic00}.  Most of the discussions
on jet-influenced star formation so far have concentrated on the global properties
of their host galaxies.  It remains to be seen whether or not GC formation can
be either inhibited or enhanced relative to field-star formation by AGN activity.

An opposite approach, that GCs can \emph{build} or at least enlarge the central SMBH,
has been promoted in other recent papers
\citep[see, e.g.][for comprehensive discussions]{cap01,cap05,cap09}.
The essential idea is that GCs in the inner region of a galaxy can be disrupted by
dynamical friction over a timescale short enough to add significantly to the stellar
population of the central bulge \citep{tre75} and thus, perhaps, to the central black hole as well.
Their quantitative estimates are based on the assumption that the radial distribution of GCs in the
galaxy was initially the same as the distribution of field-star light, and that it is now
flatter in the central regions because GCs there were destroyed.  Under this assumption
typically 30-50\% of the original GC population may have been lost this way.  Plotting
SMBH mass versus 
the \emph{calculated} mass from the supposedly ``lost'' clusters then shows a very rough
positive correlation \citep{cap01,cap09}.  Significant difficulties confront this interpretation.
It invokes a large number of long-vanished (and massive) GCs, and 
to make a major contribution to the SMBH growth, very large numbers of stars from
the disrupted GCs would have to
find their way into a tiny, AU-sized central black hole rather than just joining the
bulge stellar population.  Another
traditional objection to such a picture is that the innermost bulge population is generally much
more metal-rich \citep[reaching Solar or higher; e.g.][]{joh13} than all but a few individual GCs (which mostly range
from 1/3 to 1/200 Solar), so only a small fraction of the bulge can have been built
by cluster disruption.  Furthermore, there is no evidence that even the most massive GCs
in existence
($10^7 M_{\odot}$ and above) are sufficiently metal-rich to match the inner bulge population
\citep{har09}.  

Yet another approach would be to assume that intermediate-mass BHs already 
residing in the GCs could combine to add to the galaxy's SMBH.  Little is known about the
presence or actual masses of IMBHs though they have been proposed as seeds for SMBH 
formation \citep[see, e.g.][for a recent discussion]{jal12}.  
One of the best observational cases for an IMBH is in $\omega$ Centauri,
for which plausible evidence exists that an IMBH of 50,000 $M_{\odot}$ is present \citep{jal12}.  If we suppose that
(a) all GCs contain IMBHs, (b) their IMBH mass is directly proportional to cluster mass
with a mass ratio of 2\% like $\omega$ Cen,
and (c) all the IMBHs in the disrupted GCs add to the central SMBH, then on the average
1000 GCs (a typical value for a moderately large elliptical)
could add $\sim 5 \times 10^6 M_{\odot}$ to the SMBH.  All of these assumptions are arbitrary,
and in any case this added mass is only a few percent of the typical $\sim 10^8 M_{\odot}$ SMBH mass for
the same type of galaxy (see below).

Recently, \citet{jah11} explored the very different possibility, suggested by \citet{peng07}, that the
origin of the 
 \mbh - $M_{bulge}$ (equivalent to our \mdyna) relation results from a statistical convergence process
rather than a causal, physical one.  They used numerical simulations and merger trees to demonstrate
that hierarchical galaxy formation leads to a linear relation between SMBH and bulge mass even if 
their initial distributions were strongly uncorrelated.
The initial \mbh - $M_{bulge}$ correlation they derived from this simple assumption 
was, however, much tighter than is observed and 
had a different slope.  By adding other features to their simulations (including star formation, black 
hole accretion, and disk to bulge mass conversion) they produced correlations much more similar to
observed data and concluded that the relation ``is produced naturally by the merger-driven assembly 
of bulge and BH mass, and without any coupling of SF and BH mass growth per individual galaxy''.   

Our conclusion for the present is that the likeliest chance to establish a causal link
between the GCS and SMBH (if there is one) would be through AGN feedback and its influence on 
star and GC formation.
Much ground remains to be explored in this direction through modelling.
The possibility of a noncausal, statistical origin also remains competitive.

\begin{table*}
\caption{Selected Global Properties of Galaxies}
\label{alldata}
\begin{tabular}{lccccccccc}
\hline
Galaxy & Type & d (Mpc) & $N_{GC} (\pm)$ & log $M_{BH}/M_{\odot} (\pm)$ & 
$\sigma_e (\pm)$ (km/s) & $R_{e}$ (kpc) & log $(M_{dyn}/M_{\odot})$ & References \\
\hline
 MilkyWay & Sbc &  ~~0.0 &   160  (10) &   6.61(+0.04,-0.04) & 105(20) &  ~0.70 &  ~9.73 & 2,3,4 \\
 NGC221 &    E2  & ~~0.8 &     1   (9) &   6.46(+0.08,-0.10) &  75( 3) &  ~0.15 &  ~8.77 & 2,5,6,7,8 \\
 NGC224  &  Sb  &  ~~0.7 &   450 (100) &   8.18(+0.20,-0.10) & 160( 8) &  ~1.06 &  10.28 & 9,10,11,12,13,14 \\
 NGC253 &   SBc &  ~~3.4 &    20  (10) &   7.00(+0.30,-0.30) & 109( 5) &  ~0.31 &  ~9.41 & 15,16 17,18 \\
 NGC821 &   E4  &  ~24.1 &   320  (40) &   8.23(+0.16,-0.25) & 209(10) &  ~5.32 &  11.20 & 19,20 \\
 NGC1023 &  SB0 &  ~11.4 &   221 (100) &   7.64(+0.04,-0.04) & 205(10) &  ~1.98 &  10.76 & 21,22 \\
 NGC1316 &  E   &  ~21.5 &   647 (100) &   8.24(+0.17,-0.23) & 226( 9) &  ~3.75 &  11.12 & 23,24,25 \\
 NGC1332 &  S0  &  ~22.9 &  1000 (500) &   9.17(+0.06,-0.06) & 328(16) &  ~0.93 &  10.82 & 1,26,27 \\
 NGC1399 &  E1  &  ~20.0 &  6625(1180) &   8.69(+0.06,-0.07) & 337(16) &  ~4.11 &  11.51 & 28,29,30,31 \\
 NGC2778 &  E2  &  ~22.9 &    50  (30) &   7.15(+0.30,-0.30) & 175( 8) &  ~1.83 &  10.59 & 2,32,20 \\
 NGC3031 &  Sb  &  ~~3.6 &   172 (100) &   7.84(+0.11,-0.06) & 143(16) &  ~1.74 &  10.39 & 33,34 \\
 NGC3115 &  S0  &  ~~9.7 &   550 (150) &   8.96(+0.19,-0.15) & 230(11) &  ~1.70 &  10.79 & 35,36 \\ 
 NGC3377 &  E5  &  ~11.2 &   266  (66) &   8.25(+0.19,-0.29) & 145( 7) &  ~1.83 &  10.43 & 26,01 \\
 NGC3379 &  E1  &  ~10.6 &   270  (68) &   8.04(+0.22,-0.26) & 206(10) &  ~1.80 &  10.73 & 37,38 \\
 NGC3384 &  SB0 &  ~11.6 &   128 (100) &   7.04(+0.16,-0.26) & 143( 7) &  ~0.73 &  10.02 & 37,20 \\
 NGC3414 &  S0  &  ~25.2 &   400 (200) &   9.40(+0.06,-0.08) & 205(11) &  ~4.1  &  11.08 & 26,57,69 \\
 NGC3585 &  S0  &  ~20.0 &   300 (100) &   8.51(+0.15,-0.13) & 213(10) &  ~3.83 &  11.08 & 1,39,40 \\
 NGC3607 &  E   &  ~22.8 &   600 (200) &   8.15(+0.10,-0.15) & 229(11) &  ~7.25 &  11.42 & 1,26,40 \\
 NGC3608 &  E2  &  ~22.9 &   450 (200) &   8.67(+0.09,-0.10) & 182( 9) &  ~3.91 &  10.96 & 1,26,20\\
 NGC3842 &  E   &  ~98.4 &  9850(3900) &   9.99(+0.11,-0.13) & 270(14) &  17.8  &  11.96 & 1,41,42 \\  
 NGC4261 &  E2  &  ~31.6 &   530 (100) &   8.71(+0.08,-0.09) & 315(15) &  ~5.92 &  11.61 & 2,43,44 \\
 NGC4291 &  E2  &  ~26.2 &  1200 (600) &   8.98(+0.10,-0.18) & 242(12) &  ~1.87 &  10.88 & 1,26,20 \\
 NGC4350 &  S0  &  ~15.4 &   196  (60) &   8.74(+0.10,-0.10) & 175( 5) &  ~1.37 &  10.47 & 45,46 \\
 NGC4374 &  E1  &  ~18.4 &  4300(1200) &   8.96(+0.05,-0.04) & 296(14) &  ~4.87 &  11.47 & 45,47 \\
 NGC4382 &  E2  &  ~18.4 &  1100 (181) &   7.81(upper limit) & 182( 5) &  ~6.37 &  11.17 & 45,48 \\
 NGC4459 &  E2  &  ~16.1 &   218  (28) &   7.84(+0.08,-0.08) & 167( 8) &  ~1.16 &  10.35 & 45,49 \\
 NGC4472 &  E2  &  ~16.3 &  7800 (850) &   9.26(+0.12,-0.18) & 310(10) &  ~8.22 &  11.74 & 45,50 \\
 NGC4473 &  E5  &  ~15.7 &   376  (97) &   7.95(+0.19,-0.29) & 190( 9) &  ~1.90 &  10.67 & 45,20 \\
 NGC4486 &  E0  &  ~16.1 & 13300(2000) &   9.77(+0.03,-0.03) & 324(16) &  ~7.80 &  11.76 & 45,51,52 \\
 NGC4486A & E2  &  ~15.6 &    11  (12) &   7.15(+0.13,-0.11) & 111( 5) &  ~0.53 &  ~9.66 & 45,53 \\
 NGC4486B & E1  &  ~16.3 &     4  (11) &   9.00(upper limit) & 185( 9) &  ~0.24 &  ~9.76 & 45,54 \\
 NGC4552 &  E1  &  ~15.3 &  1200 (250) &   8.68(+0.04,-0.05) & 233(11) &  ~2.20 &  10.92 & 45,51,55 \\
 NGC4564 &  S0  &  ~15.0 &   213  (31) &   7.92(+0.10,-0.14) & 162( 8) &  ~1.58 &  10.46 & 45,20 \\
 NGC4594 &  Sa  &  ~~9.8 &  1900 (300) &   8.72(+0.30,-0.57) & 240(12) &  ~2.53 &  11.01 & 19,56 \\
 NGC4621 &  E4  &  ~18.3 &   800 (355) &   8.60(+0.06,-0.07) & 225(11) &  ~4.12 &  11.16 & 45,57 \\
 NGC4649 &  E2  &  ~16.8 &  4745(1100) &   9.63(+0.09,-0.11) & 385(19) &  ~7.30 &  11.88 & 45,20 \\
 NGC4697 &  E6  &  ~11.7 &   229  (50) &   8.28(+0.08,-0.10) & 177( 8) &  ~4.27 &  10.97 & 45,20 \\
 NGC4889 &  E   &  103.2 & 11000(1340) &  10.32(+0.25,-0.54) & 347(17) &  25.1  &  12.32 & 58,42 \\
 NGC5128 &  E0p &  ~~3.8 &  1300 (300) &   7.71(+0.13,-0.19) & 150( 7) &  ~5.60 &  10.94 & 59,55,60,61 \\
 NGC5813 &  E2  &  ~32.2 &  1650 (400) &   8.84(+0.07,-0.07) & 239(12) &  ~7.60 &  11.48 & 62,57 \\
 NGC5845 &  E3  &  ~25.9 &   140  (70) &   8.69(+0.12,-0.16) & 234(11) &  ~0.51 &  10.29 & 1,26,20 \\
 NGC5846 &  E0  &  ~24.9 &  4700(1200) &   9.04(+0.07,-0.09) & 237(12) &  ~9.97 &  11.59 & 63,57 \\
 NGC6086 &  E   &  132.0 &  4584(1362) &   9.56(+0.16,-0.16) & 318(16) &  20.30 &  12.16 & 1,64,42 \\
 NGC7332 &  S0  &  ~23.0 &   175  (15) &   7.11(+0.17,-0.21) & 116( 6) &  ~1.2  &  10.05 & 70,57  \\
 NGC7457 &  S0  &  ~13.2 &   178  (75) &   7.00(+0.20,-0.40) &  67( 3) &  ~0.90 &  ~9.45 & 65,20 \\
 NGC7768 &  E   &  112.8 &  3000(1300) &   9.11(+0.15,-0.27) & 257(13) &  12.6  &  11.76 & 64,71,42 \\
 IC1459 &   E4  &  ~29.2 &  2000 (800) &   9.45(+0.14,-0.25) & 340(17) &  ~5.47 &  11.64 & 1,26,55 \\
 IC4296 &   BCG &  ~50.8 &  6400(1600) &   9.13(+0.06,-0.07) & 322(12) &  10.20 &  11.86 & 1,66,67 \\
 A2052  &   BCG &  141.0 & 27000(9000) &   9.66(upper limit) & 233(11) &  38.30 &  12.16 & 68,67 \\
\hline
\end{tabular}

Sources: (1) This paper, (2) \citet{hh}, (3) \citet{g11}, (4) \citet{g12}, (5) \citet{l1},
(6) \citet{v1}, (7) \citet{gr96}, (8) \citet{fi10}, (9) \citet{b6}, (10) \citet{b7}, (11) \citet{f06},
(12) \citet{br04}, (13) \citet{mc05}, (14) \citet{sa09}, (15) \citet{ol04}, (16) \citet{ro06}, 
(17) \citet{da09}, (18) \citet{re05}, (19) \citet{s1}, (20) \citet{sg11}, (21) \citet{l2}, u
(22) \citet{b5}, (23) \citet{v3}, (24) \citet{ri11}, (25) \citet{n2}, (26) \citet{k1}, 
(27) \citet{ru11}, (28) \citet{b3}, (29) \citet{d1}, (30) \citet{h3}, (31) \citet{g5}, 
(32) \citet{g1}, (33) \citet{s4}, (34) \citet{d2}, (35) \citet{h6}, (36) \citet{e1}, 
(37) \citet{h2}, (38) \citet{g13}, (39) \citet{p3}, (40) \citet{g14}, (41) \citet{bu92}, 
(42) \citet{mcc11}, (43) \citet{g9}, (44) \citet{f1}, (45) \citet{p1}, (46) \citet{p4}, 
(47) \citet{b4}, (48) \citet{g10}, (49) \citet{s3}, (50) \citet{s2}, (51) \citet{t2}, 
(52) \citet{ge11}, (52) \citet{n1}, (54) \citet{ko97}, (55) \citet{c1}, (56) \citet{k2}, 
(57) \citet{g7}, (58) \citet{ha09}, (59) \citet{h1}, (60) \citet{neu10}, (61) \citet{ha10}, 
(62) \citet{h4}, (63) \citet{f3}, (64) \citet{bl97}, (65) \citet{c2}, (66) \citet{ok02}, 
(67) \citet{db09}, (68) \citet{ha95}, (69) \citet{hu08}, (70) \citet{fo01}, (71) \citet{jo04}
\end{table*}

\section[]{The Data}

In a parallel study, HHA13 have compiled a catalog of 422 galaxies of all types with
 \ngc data published to the present time.  Among these there are now 49 galaxies with
measurements of both \mbh and \ngca: 34 ellipticals (E), 10 ``lenticulars'' (S0), 
and 5 spirals (S).  As discussed by HH, the dominance of E galaxies in our sample is partly because 
more central black hole masses are known for ellipticals, but partly because the best S0 and S 
galaxies in which to determine \mbh are face-on or nearly so.  By contrast, \ngc studies 
in disk galaxies are best done for edge-on systems, minimizing the disk contamination
and background light.
Thus for disk galaxies, the globular cluster studies and SMBH studies in the literature
have traditionally had little overlap.  The situation for the E's is much better.

As hinted above, a primary issue in interpreting the correlations that we derive
in the discussion below is that the 
dispersions (rms scatters) found around the best-fit relations 
are consistently larger than expected from the formal errors quoted in the literature for
the individual measured parameters.  Before proceeding with our correlation solutions, therefore, we
(1) discuss just what typical ranges of values have been found 
in different studies for each key quantity, and 
(2) develop as self-consistent a database as possible for our sample.  
We discuss in turn the data for galaxy distance, effective radius, velocity dispersion,
\ngca, and bulge mass and \mbha.

\subsection[]{Distance}
As noted above, we are primarily interested in galaxies where both \mbh and \ngc
are measured and our final adopted list is in Table 1.
Not surprisingly, these are in cosmologically nearby space and so 
also have published measurements of \sige and \re from which their dynamical
masses $M_{dyn}$ can be directly calculated once the distance is known.  However,
because their distances range from less than 1Mpc to more than 100Mpc the best available distances 
are themselves based on a variety of techniques.  As a result, it is not feasible to apply any single 
distance method to the whole sample.  Because \rea, $M_{dyn}$, and \mbh all scale with distance, 
we chose first to assess distance values for each galaxy individually.

The distances quoted in Table 1 are based on three methods: stellar standard candles, surface 
brightness fluctuations (SBF), and redshift (i.e. Hubble's law).  For the six galaxies nearer than 4Mpc it is 
possible to resolve individual stars and use standard candles such as RR Lyrae stars, Cepheids, 
and the luminosity of the tip of the red giant branch, with resulting  
uncertainties of $\sim$5$\%$.  SBF distances are available for the great majority of our sample
and we use them for the galaxies between 4 Mpc and 50 Mpc; formal 
SBF uncertainties range from $\sim$ 4-15$\%$. 
For five galaxies beyond $\sim$50 Mpc, we use the galaxy's redshift 
relative to the cosmic microwave background (CMB) and $H_o$ = 70$\pm$2 km/s/Mpc.  
Formal uncertainties for the most distant galaxies in our sample are $\sim$7$\%$.

Since publication of the first major SBF sample by \citet{ton01}, the calibration has been 
revised several times.  \citet{jen03} used Hubble Space Telescope (HST) $\it IR$ imaging
to study 65 galaxies while \citet{bla09} recalibrated distances for 134 galaxies in the Virgo 
and Fornax clusters, again with HST.  Unfortunately for our purposes, neither of these had 
more than a 50$\%$ overlap with our sample.  For the galaxies in \citet{ton01} in common between \citet{jen03} 
or \citet{bla09} we find a mean difference of $\Delta(m-M)_0(T-B/J) = 0.16\pm 0.10$.
Given the lack of compelling evidence to support any large adjustment to the 
\citet{ton01} values, which overlap with $\sim$85$\%$ of our sample, we have chosen 
to use the 2001 distances here. 

\citet{mcc11} use redshift-based distances with $H_o$ = 70 
 for all of the galaxies in their compilation 
and, given the mix of distance methods we have applied, it is useful to compare the two datasets
for the 39 galaxies in common.  In Figure 1 we compare their redshift distances with our 
resolved-star and SBF values.  As seen in the top panel, the distance difference ($d-d_z)$ is, on
average, less than the formal error in the stellar/SBF distances.  In addition (lower panel), 
the redshift distances are systematically larger, although within the errorbars shown.

\begin{figure}
 \includegraphics[width=0.45\textwidth]{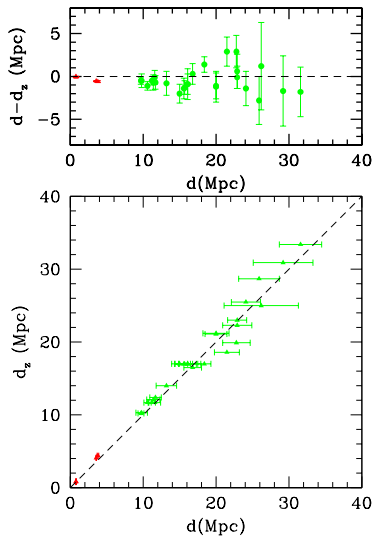}
 \caption{Comparison of redshift-based distances with those from stellar standard candles
 (red) and surface brightness fluctuations (green). In the upper panel the difference $d-d_z$
is plotted against distances based on stellar indicators such as
resolved stars and SBF.  In the lower panel 
we plot $d_z$, the redshift-based distance, against the stellar-based value.}
\end{figure}

For most of the following discussion we will use the mainly non-redshift distances given
in Table 1.  

\subsection[]{Effective Radius}
The effective radius of a galaxy (\rea) is the radius of the isophote containing half of 
the total luminosity (cf. \citet{bin08} p. 21).  The definition is simple in concept
but much less simple to measure, because it requires discriminating data at both the inner galaxy,
to determine the central profile, and at large galactocentric radii, to determine the outer 
limits of the galaxy profile.  In addition, the results can be wavelength dependent due 
primarily to the relative contribution of young stars, old stars, and dust.  
Consequently, the effective radius values in the literature can differ considerably.  

Since \re is an essential piece in determining a galaxy's \mbha, \mdyna, and
location in the fundamental plane we wanted to understand its uncertainty.  In Figure 2
are plotted the values of \re that we found for our galaxy sample.  The values 
plotted were based on data from the litarature adjusted to our distances 
as given in Table 1; adoption of the redshift-based distances would change the details but 
not the overall picture.  It is important to note here that none of the \re catalogues
we used, except those from the 2MASS Extended Sources Catalogue \citep{sk06}, 
contained all of our galaxies.  In addition, the methods used are also 
very different.  As seen in Figure 2, the range of \re for a given galaxy can be almost a factor of
two.  Ideally the value of \re for each galaxy would be based on detailed comparisons of 
available sources that take into account their strengths and limitations.  But, because of 
the range of methods, limiting magnitudes, wavelength of the observations, and differing overlap,
we decided to simply use the \re given in the \mbh determination paper for each galaxy.

Note that the data shown are not meant to represent a complete list of all \re data in the literature.  
Rather, this discussion is intended to show the range of values found for a given galaxy and act as a 
reminder that this quantity can be difficult to measure consistently.

\begin{figure}
 \includegraphics[width=0.45\textwidth]{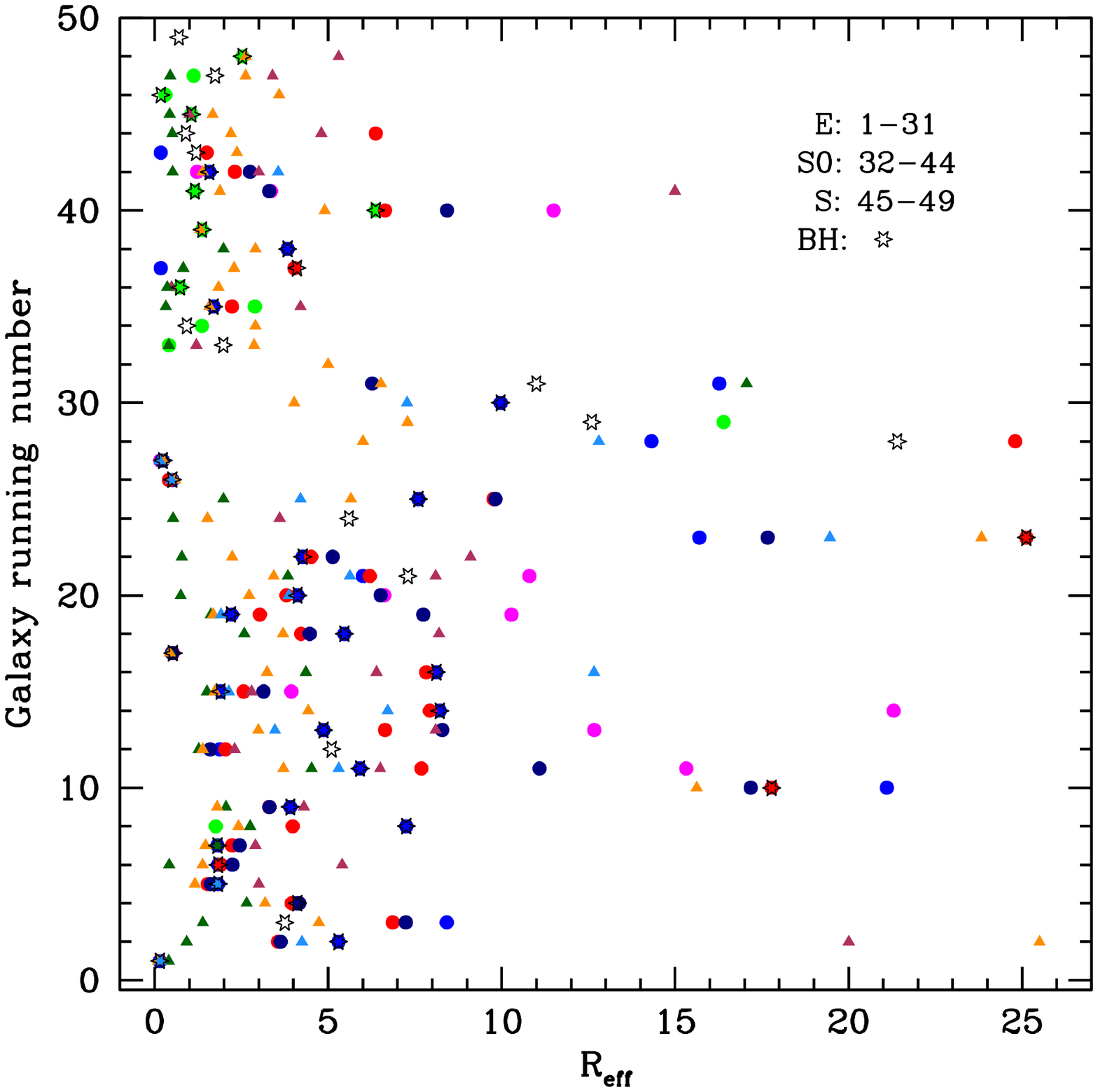}
 \caption{Values of \re from various sources are plotted against running galaxy number.  Filled circles are
blue: \citet{fa89}, green; \citet{b3}, red: \citet{l07}, magenta: \citet{ko09}, navy blue: \citet{ho09}. 
Filled triangles are maroon: \citet{mh03}, dark green: \citet{hu08}, dark orange: (2MASS), dodger blue: \citet{m5}
The black starred symbol is the value of \re used in the determination of \mbh. }
\end{figure}

\subsection[]{Velocity Dispersion}
A search of the HyperLeda \citep{pat03} database shows that published \sige values 
also range widely, by even as much as $\sim$100 km/s or more for a single galaxy.
\citet{mce95} compared results from 
different sources and tabulated velocity dispersions for more than 1500 galaxies 
based on his analysis of 85 standard galaxies.  In Figure 3 the 15 McElroy  
standard galaxies in our sample are shown in red with the rest in green.  Superimposed 
on the line for each galaxy are also the McElroy value (blue), the HyperLeda value 
(black *) and the value quoted in the \mbh paper (black x).
We note that, formally, the expected  
uncertainties in \sige are often quoted as $\leq 5\%$ \citep[cf.][]{g4}, but the scatter among
different studies for the same galaxy is obviously much larger than that.  The discordance between 
this and the data plotted in Figure 3 suggests that observational factors such as spatial 
resolution, choice of \rea, and limiting magnitude are important but difficult to correct
for in a given data set.  Another way to compare the values in the literature is shown 
in Figure 4 where we have simply plotted the \sige value adopted in the black hole mass
determination (x-axis) vs. the HyperLeda and McElroy average values (y-axis).
Interestingly, except for a few galaxies, the values are consistent within
about twice the nominal $\pm 5\%$ scatter usually adopted for \sige. 
Thus, as with \rea, our tabulated values of \sige are those used 
in the \mbh determination studies.  We believe this makes the most sense when
comparing galaxy parameters such as \sigea, \mbha, and \mdyna.

\begin{figure}
 \includegraphics[width=0.45\textwidth]{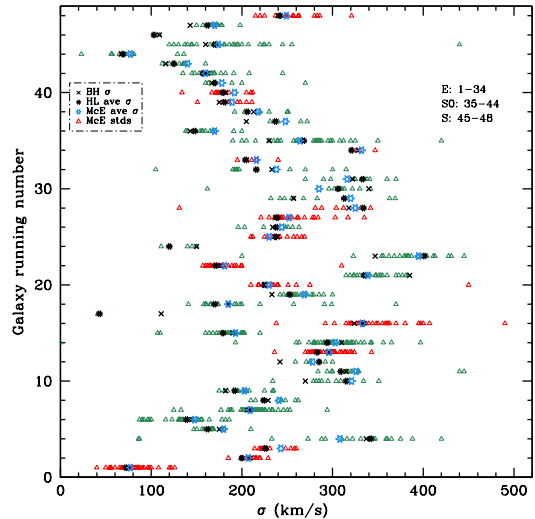}
 \caption{Plotted are velocity dispersion values from HyperLeda for the galaxies in our sample
against running galaxy number.  Open red trianlges are McElroy ``standard'' galaxies and green triangles
are the rest of the sample.  The black X represents the velocity dispersion used in the black hole 
mass determination, the black asterisk is the average value from HyperLeda, and the open blue
asterisk is the McElroy average for each galaxy.}
\end{figure}

\begin{figure}
 \includegraphics[width=0.45\textwidth]{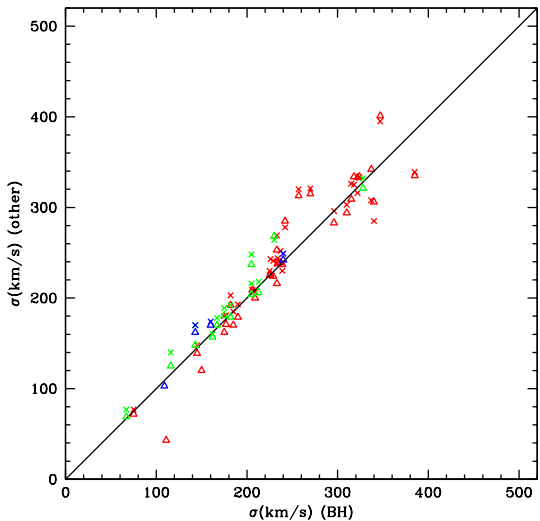}
 \caption{\sige as used in the \mbh studies is plotted on the x-axis vs. mean values for \sige from Hyperleda (open 
triangles) and McElroy (x's) on the y-axis.  E galaxies are shown in red, S0 in green and S in blue.}
\end{figure}  

\subsection{Central Black Hole Masses}
The number of galaxies with measurements of \mbh has increased significantly in the past
decade and is now approaching 100, with more to come.  Most of the data in Table 1 can be found in 
the compilations of \citet{g7}, \citet{g8}, and \citet{mcc11}, supplemented by a few
additional studies.  Generally the masses we quote are from a single source, though a small
number of galaxies have more than one determination.  In the case of NGC 1399 we have used a weighted
mean of the results from \citet{g5} and \citet{h3}.  The case of NGC 5128 is somewhat more complex.  
\citet{g8} list \mbh = $7.0 \times 10^7M\odot$ from \citet{c1} and $3 \times 10^8M\odot$ from \citet{si05}, 
but choose to plot the latter on their $M - \sigma$ relation.  We have chosen to follow the reasoning of 
\citet{neu07} and \citet{neu10}, averaging the two values they provide based on stellar and gas
dynamics.  This smaller \mbh brings NGC 5128 into better agreement with the $M - \sigma$ relation, 
but not for the \ngc - \mbha~ relation.  Finally, the black hole mass for NGC 4486B \citep{ko97} is 
an upper limit which is uncertain because of both its probable low mass and the galaxy's
asymmetric double nucleus.  We include it here for completeness but do not use it in the correlations below.

\subsection[]{Cluster Populations}
The data in Table 1 for the total numbers of globular clusters $N_{GC}$ are taken from 
the new HHA13 catalog, resulting in the addition of 12 more galaxies than used in HH.
Although this is a considerably larger sample than in the initial discussions of 
either Spitler \& Forbes (2009) or
Burkert \& Tremaine (2010), our list is still dominated by E galaxies.

The quoted uncertainties on \ngc in Table 1 differ quite a bit from one galaxy to another;
typically they may be at the level of $\pm 20$\% or so but they depend very much on the
details of the individual photometric studies from which they were drawn.  
The uncertainties are relatively low in cases where the study had both wide area
coverage and deep photometric limits
reaching past the turnover point of the GC luminosity function.  In these cases
\ngc {\sl by definition} is simply calculated as twice the total number of observed clusters brighter
than the turnover point, summed over all galactocentric radii.  
In several other cases \ngc is derived from smaller-field coverage or photometric limits
that fall short of the GCLF turnover, requiring larger extrapolations to estimate the
total population.  In some other cases such as for small satellites of nearby much larger
galaxies (M32, NGC 4486B) only rough guesses can be made.

\subsection[]{Bulge Mass}
The final quantity listed in Table 1 is \mdyna, generally referred to as the virial
bulge mass, given by

\begin{equation}\label{eqn-M_dyn}
M_{dyn}{=}k R_e \sigma_e ^2/G
\end{equation}

\noindent where we use $k$=3 instead of 8/3, following \citet{mh03}.  
In the case of \mdyna we have not quoted formal uncertainties in Table 1, 
primarily because consistent observational uncertainties are not generally availble 
for the values of \rea we have used.  However, sample calculations based on typical 
observational uncertainties for \re indicate that the resulting uncertainty in log\mdyn
is $<0.1$.  This is the value used in the statistical analyses described in section 4.


\section{Analysis}

\subsection[]{Reduced $\chi ^2$ Estimator}

Our first set of correlation solutions follows the method outlined in
\cite{t1} and \cite{novak06} and also used in HHA13.  Briefly, if we assume data pairs $(x_i, y_i)$ are
related by $y = \alpha + \beta x$ and that they have individual measurement
uncertainties $(\sigma_{x,i}, \sigma_{y,i})$, then we minimize
\begin{equation}\label{equation-chi2}
\chi^2 \, = \, \sum{ {(y_i - \alpha - \beta (x_i - \langle x \rangle)^2} \over {(\sigma_{y,i}^2 + \epsilon_y^2)
    + \beta^2 (\sigma_{x,i}^2 + \epsilon_x^2) } } \, .
\end{equation}
Here, $\epsilon_x$ and $\epsilon_y$ represent any additional intrinsic or ``cosmic"
scatter over and above the nominal measurement uncertainties.  If the quoted 
($\sigma_x, \sigma_y$) values genuinely represent all the scatter present in the data,
then we would have $\epsilon_x = \epsilon_y = 0$; but if additional cosmic scatter exists,
then the $\epsilon$'s can be adjusted upward from zero until we obtain a reduced
$\chi_{\nu}^2 = 1$ from the fit.  Of course, if we have only two parameters to 
correlate, it may be impossible to decide whether additional intrinsic scatter
is in $x$, or in $y$, or in both.  However, if we have several parameters to work with, it may
be possible to find a self-consistent set of $\epsilon$'s for them all from
the several 2-parameter correlations they allow.

Our first set of solutions, based on the data in Table 1 and listed in Table \ref{Tab2} below, assumes for each pair of
parameters that $\epsilon_x = 0$ and all the additional scatter (if any) is in $y$.
That is, $\epsilon_y$ is a free parameter adjusted to make 
$\chi_{\nu}^2 = 1$.  As is discussed at greater length in previous studies of
this type \citep[][among others]{t1, novak06}, including the $\epsilon$ factor
directly changes the relative weightings of the datapoints and thus noticeably
influences the best-fitting slope $\beta$.  The larger the adopted $\epsilon$, the
more uniform the total weights of the datapoints become.
The astrophysical importance is clearly that {\sl if} we need to
invoke nonzero $\epsilon$, then either the measurement uncertainties have
been underestimated, or the galaxies do have an inbuilt variance that higher
precision data would not remove.

In Table \ref{Tab2} the successive columns give (1) the pair of correlated quantities $(y, x)$;
(2,3) the galaxy types included in the fit and the number of individual datapoints;
(4, 5, 6) the zeropoint $\alpha$ and slope $\beta$ of the fit, with internal uncertainties,
and the sample mean $\langle x \rangle$;
(6) the residual rms scatter $\sigma_y$ about the best-fit line; and (7) the added
intrinsic scatter $\epsilon_y$ needed to make the total goodness of fit $\chi_{\nu}^2$ 
equal unity.  The footnotes list individual galaxies excluded from each fit, either because of 
missing or uncertain data or because they are extreme outliers (these are more evident
in the Figures shown below).

\begin{table*}
\caption{Correlation Solutions from $\chi^2$ Minimization}
\label{Tab2}
\begin{tabular}{llccccccl}
\hline
Solution For: & Sample & N & $\alpha$ & $\beta$ & $\langle x \rangle $ & RMS $\sigma_y$ & $\epsilon_y$ & Note \\
\hline
log $N_{GC}$ vs log $\sigma_e$ & E & 29 & $2.825 \pm 0.071$ & $4.120 \pm 0.540$ & 2.30 & 0.383 & 0.37 & 6 \\
 & All & 43 & $2.821 \pm 0.055$ & $3.720 \pm 0.382$ & 2.30 & 0.362 & 0.35 & 6 \\
log $N_{GC}$ vs log $M_{dyn}$ & E & 31 & $2.810 \pm 0.053$ & $1.130 \pm 0.098$ & 11.0 & 0.295 & 0.23 & 1\\
 & All & 46 & $2.932 \pm 0.047$ & $0.870 \pm 0.068$ & 11.0 & 0.320 & 0.26 & 1\\
log $N_{GC}$ vs log \mbh & E & 29 & $2.952 \pm 0.059$ & $0.840 \pm 0.080$ & 8.5 & 0.318 & 0.30 & 3 \\
 & E+S & 33 & $2.968 \pm 0.053$ & $0.830 \pm 0.069$ & 8.5 & 0.302 & 0.27 & 4 \\
 & All & 41 & $2.968 \pm 0.050$ & $0.750 \pm 0.047$ & 8.5 & 0.323 & 0.30 & 5 \\
log \mbh vs log $\sigma_e$ & E & 34 & $8.413 \pm 0.080$ & $4.730 \pm 0.539$ & 2.30 & 0.469 & 0.44 & \\
 & All & 49 & $8.412 \pm 0.067$ & $4.610 \pm 0.403$ & 2.30 & 0.466 & 0.44 \\
log \mbh vs log $M_{dyn}$ & E & 29 & $8.497 \pm 0.060$ & $1.010 \pm 0.083$ & 11.0 & 0.323 & 0.26 & 2 \\
 & All & 44 & $8.537 \pm 0.059$ & $1.050 \pm 0.077$ & 11.0 & 0.393 & 0.34 \\
log \mbh vs log $N_{GC}$ & E   & 29 & $8.33  \pm 0.064$ & $0.930 \pm 0.097$ & 2.70 & 0.344 & 0.31 & 3 \\
 & E+S & 33 & $8.273 \pm 0.059$ & $0.980 \pm 0.090$ & 2.70 & 0.340 & 0.30  & 4 \\
 & All & 41 & $8.221 \pm 0.062$ & $1.070 \pm 0.092$ & 2.70 & 0.398 & 0.35 & 5 \\
\hline
\end{tabular}

Notes: (1) Excludes M32, NGC 4486A, NGC 4486B \\
(2) Excludes NGC 4486B, NGC 2778, NGC 4382, NGC 5128, and NGC 5845 \\
(3) Excludes M32, NGC 4486A, NGC 4486B, NGC 5128, NGC 5845 \\
(4) Excludes M32, NGC 4486A, NGC 4486B, NGC 5128, NGC 5845, Milky Way \\
(5) Excludes M32, NGC 4486A, NGC 4486B, NGC 5128, NGC 5845, Milky Way, NGC 3414, NGC 4350 \\
(6) Excludes M32, NGC 4486A, NGC 4486B, NGC 2778, NGC 7457, and A2052 \\
\end{table*}

\subsubsection[]{\ngc and \mbh Versus Velocity Dispersion}

The solutions in Table \ref{Tab2} exclude M32 (NGC 221), NGC 4486A, and NGC 4486B
 since the
globular cluster numbers in those three dwarfs are small and extremely uncertain.  However, 
all galaxies in Table 1 are plotted  in Figures 5 and 6, which also show the solutions superimposed.
The E-galaxy solution from 29 galaxies (including the outliers NGC 5128 and NGC 5845), is
$$ {\rm log} N_{GC} \, = \, (2.825 \pm 0.071) + (4.120 \pm 0.540) ({\rm log} \sigma_e - 2.3)  $$
and required invoking an extra dispersion $\epsilon_y = 0.37$ to achieve $\chi_{\nu}^2 = 1$
and has a resulting rms scatter of $\pm 0.383$ dex.  
The much larger database analyzed by HHA13 shows that the $N_{GC}$ vs. $\sigma_e$ correlation
has similarly large scatter and in addition is highly nonlinear (see their Figure 7).
For the SMBH masses, we also find for 34 E galaxies
$$ {\rm log} M_{\bullet} \, = \, (8.413 \pm 0.080) + (4.730 \pm 0.539) ({\rm log} \sigma_e - 2.3) $$ with
$\epsilon_y = 0.44$ and for $\chi_{\nu}^2 = 1$ a scatter of $\pm 0.44$ dex.
For comparison, in their original studies 
\citet{f4} found a slope of $4.8 \pm 0.5$, \citet{g13} $3.75 \pm 0.3$, 
and slopes in the range $\simeq 4 - 5$ have been commonly quoted in later papers.
We note that the absolute uncertainty in slope for 
this correlation is by far the largest for any we examined, though
the \emph{relative} uncertainty of 12\% is similar to the others.

\begin{figure}
 \includegraphics[width=0.45\textwidth]{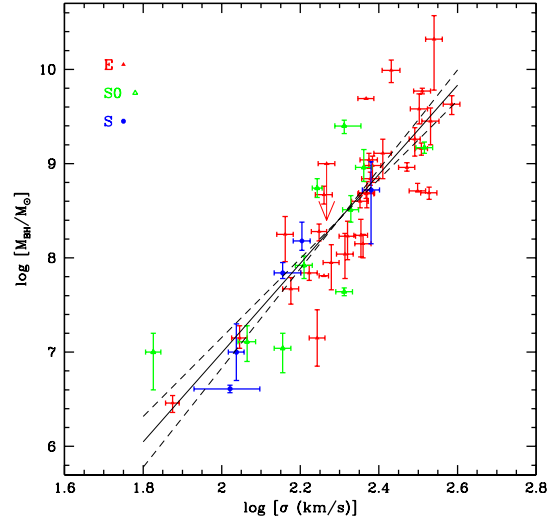}
 \caption{ Plot of \mbh vs. \sige for all 49 galaxies in our sample.  E galaxies are  
in red filled triangles; S0 are green open triangles and S are filled blue circles.  Overplotted is the 
$\chi^2$ minimization solution for all 34 E galaxies.}
\end{figure}

\begin{figure}
 \includegraphics[width=0.45\textwidth]{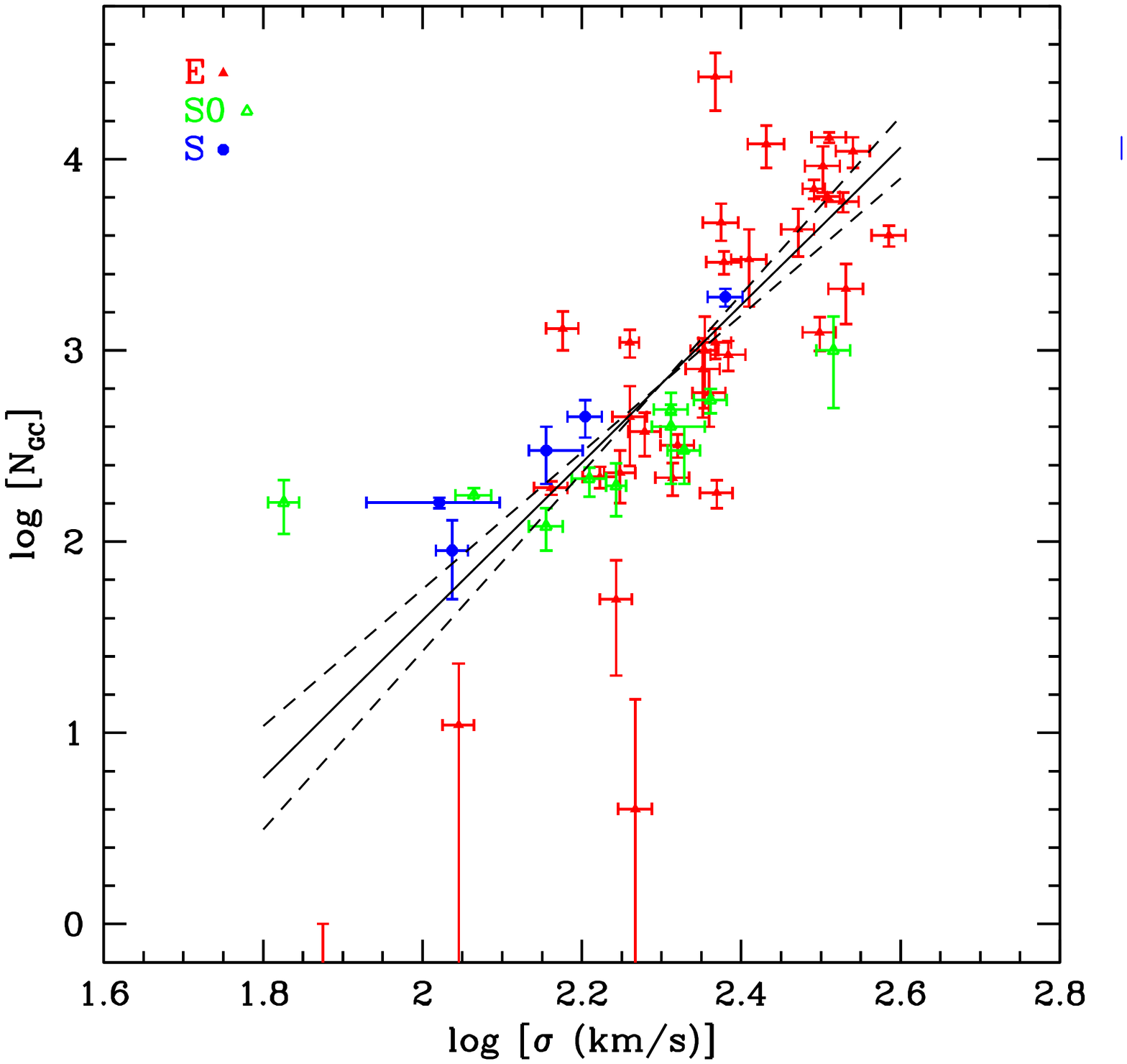}
 \caption{ Plot of \ngc vs. \sige for all 49 galaxies in our sample; symbols are as in Figure 5.   
Overplotted is the 
$\chi^2$ minimization solution for 29 E galaxies.}
\end{figure}

\begin{figure}
 \includegraphics[width=0.45\textwidth]{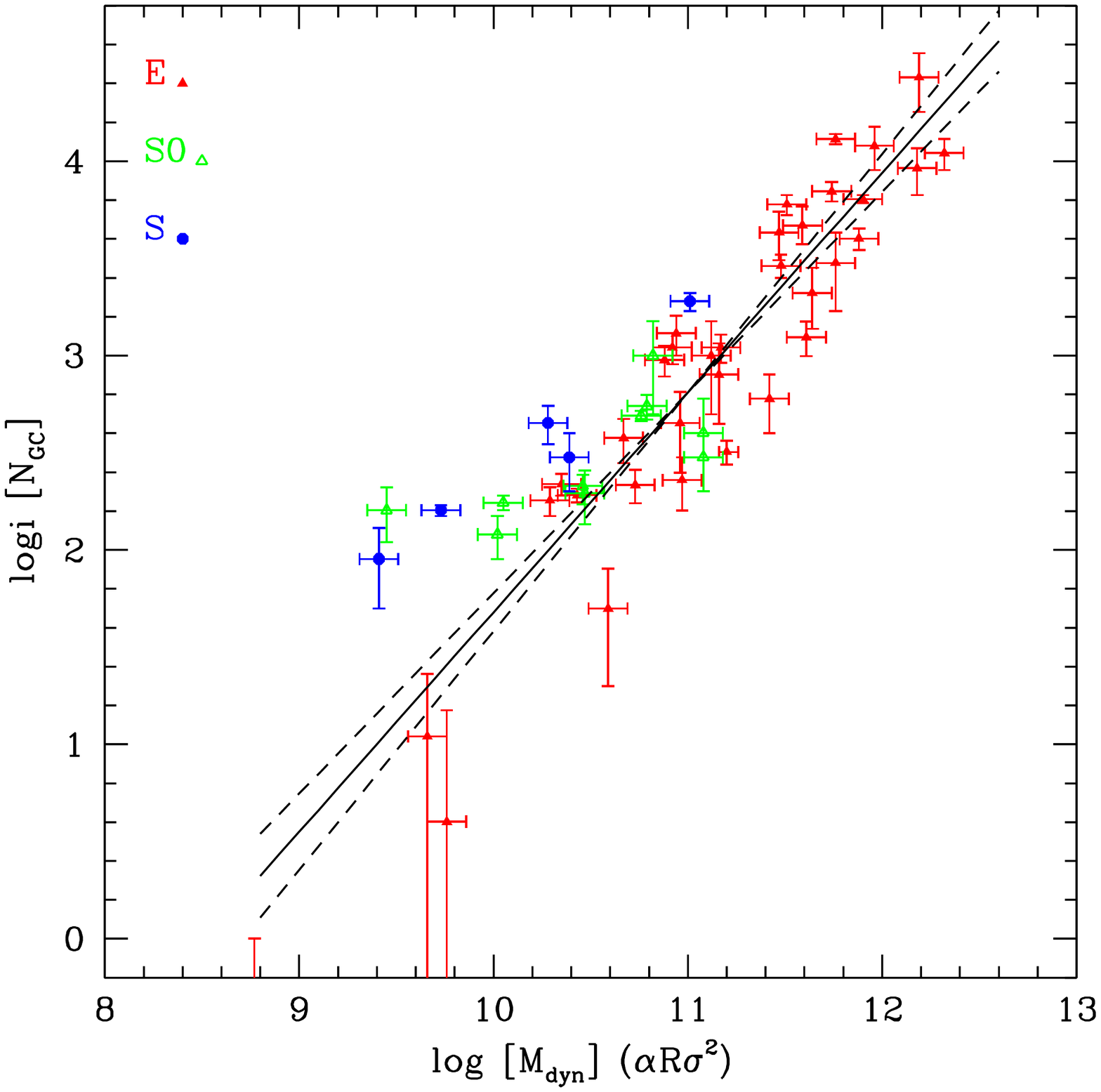}
 \caption{Plot of \ngc vs. \mdyn for all 49 galaxies in our sample; symbols are as in Figure 5.    
 Overplotted is the 
$\chi^2$ minimization solution for all 34 E galaxies.}
\end{figure}

\begin{figure}
 \includegraphics[width=0.45\textwidth]{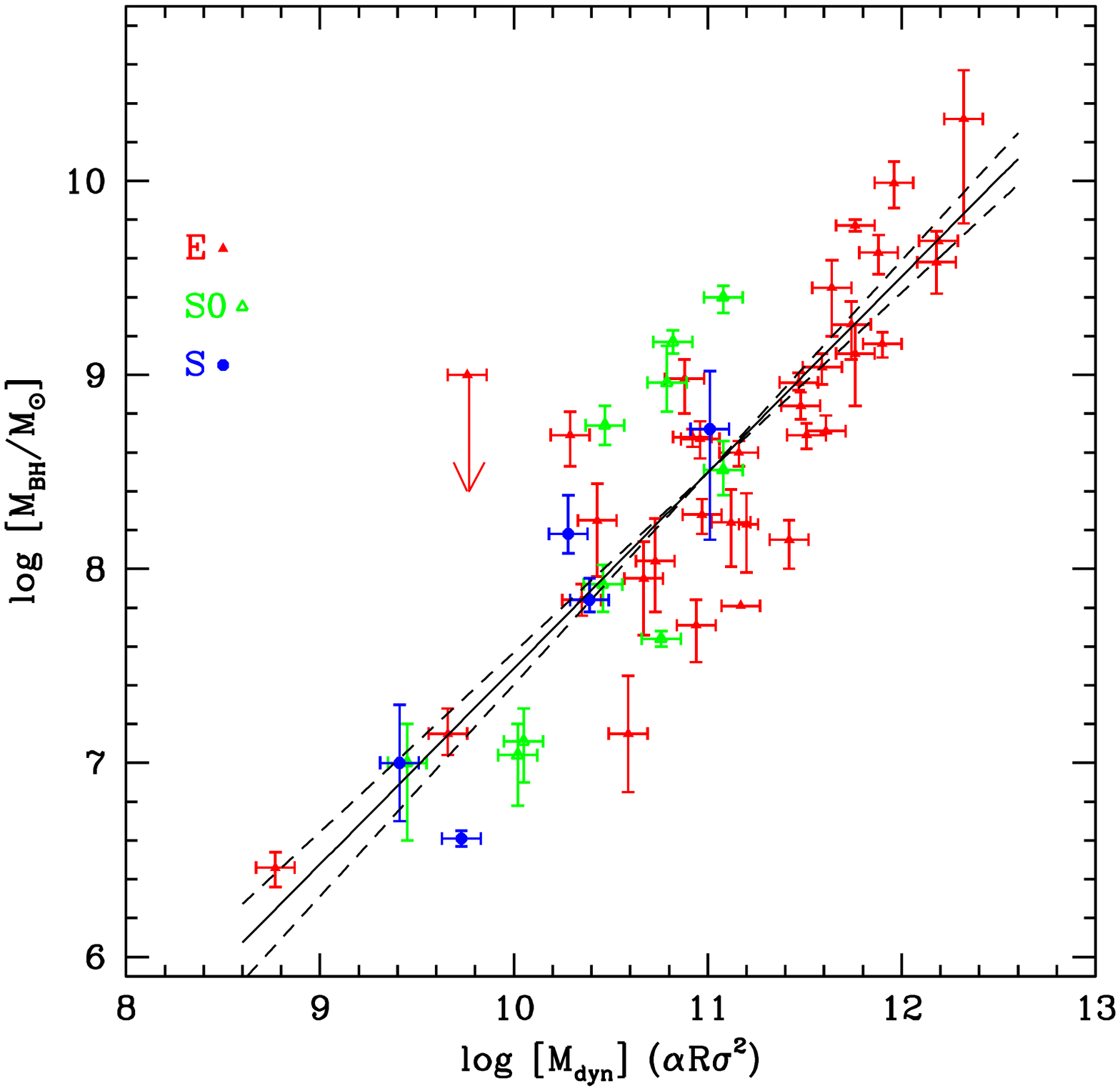}
 \caption{ Plot of \mbh vs. \mdyn for all 49 galaxies in our sample; symbols are as in Figure 5.  
 Overplotted is the 
$\chi^2$ minimization solution for 29 E galaxies.}
\end{figure}

\subsubsection[]{\ngc and \mbh Versus Dynamical Mass}

Again, the solutions in Table \ref{Tab2} exclude M32 (NGC 221), NGC 4486A, and NGC 4486B since the
globular cluster numbers in those three dwarfs are small and extremely uncertain.  
The E-galaxy solution from 31 galaxies,
$$ {\rm log} N_{GC} \, = \, (2.810 \pm 0.053) + (1.130 \pm 0.098) ({\rm log} (M_{dyn}/M_{\odot}) - 11.0)  $$
required invoking an extra dispersion $\epsilon_y = 0.23$ to achieve $\chi_{\nu}^2 = 1$
and has a resulting rms scatter of $\pm 0.295$ dex.  For comparison, the
solution found by HHA13 from 139 ellipticals with measured GCSs \emph{excluding dwarfs} was
$ N_{GC} \sim M_{dyn}^{1.035 \pm 0.033}$
with $\epsilon_y = 0.28$ and rms scatter $\pm 0.32$ dex.  These two solutions are not significantly
different, though the smaller and more highly selected sample used in this paper has a slightly
smaller dispersion.  Generally, we confirm HHA13 in that the total GC population of a
(non-dwarf) galaxy is almost directly proportional to galaxy mass.  The solution for these parameters 
is superimposed on the plot in Figure 7.

\begin{figure}
 \includegraphics[width=0.45\textwidth]{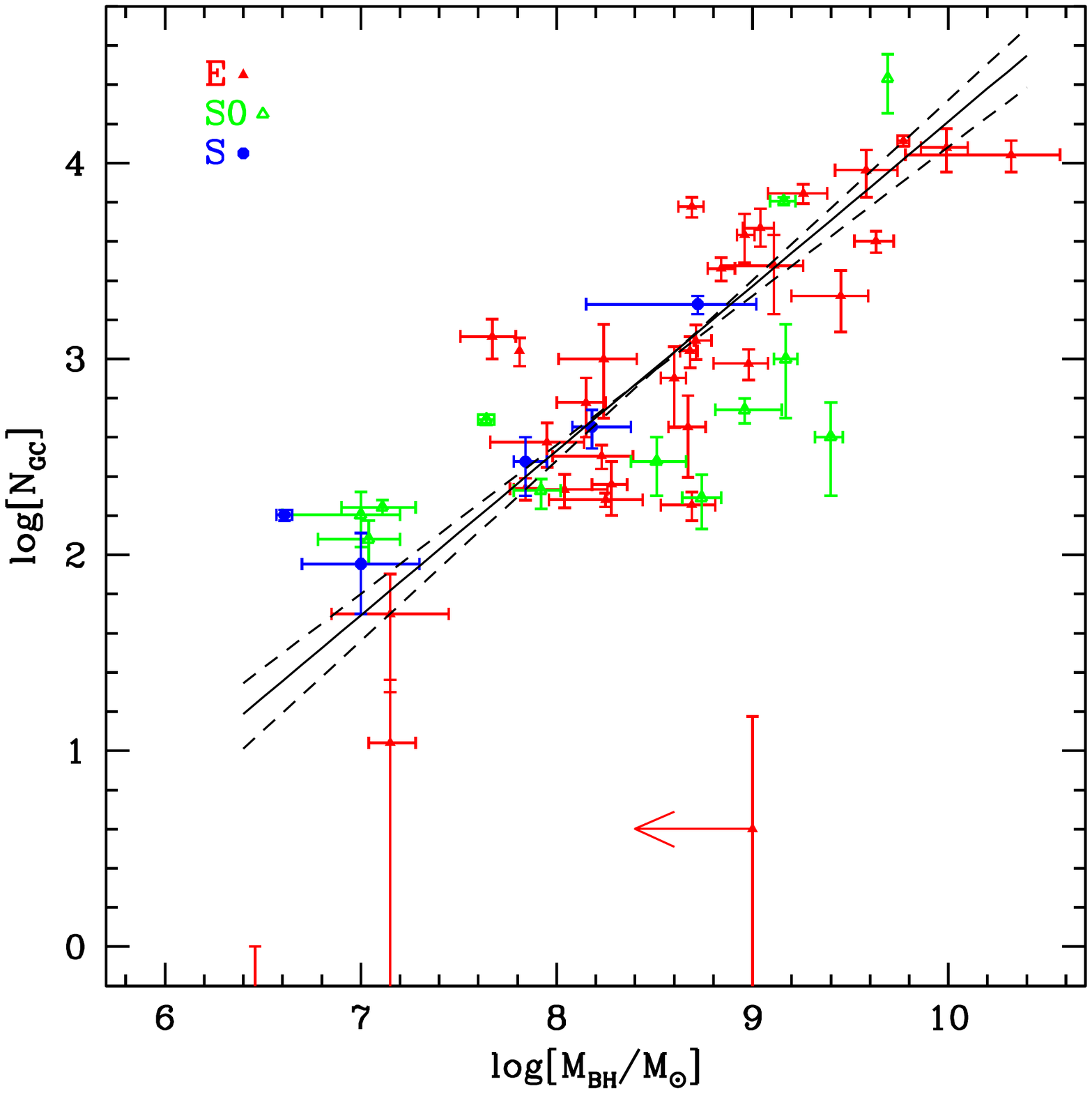}
 \caption{Plot of \ngc vs. \mbh for all 49 galaxies in our sample; symbols are as in Figure 5.    
 Overplotted is the 
$\chi^2$ minimization solution for 29 E galaxies.}
\end{figure}

\begin{figure}
 \includegraphics[width=0.45\textwidth]{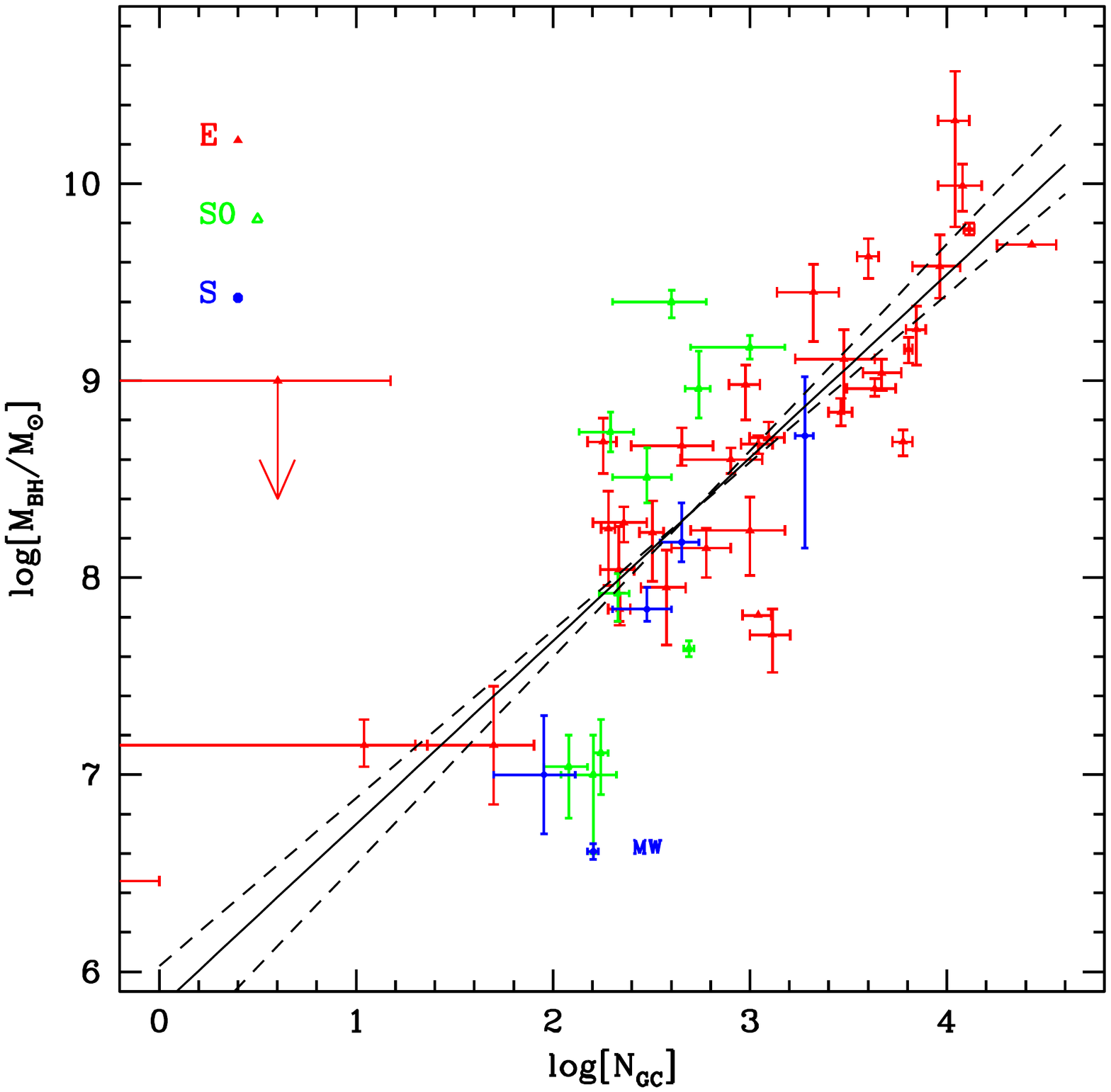}
 \caption{Plot of \mbh vs. \ngc for all 49 galaxies in our sample; symbols are as in Figure 5.    
 Overplotted is the 
$\chi^2$ minimization solution for 29 E galaxies.}
\end{figure}

For \mbh the solutions for either 29 E galaxies or 44 galaxies of all types are
not significantly different.  The larger sample (Table 2) gives
$$ {\rm log} M_{\bullet} \, = \, (8.537 \pm 0.059) + (1.050 \pm 0.077) ({\rm log} (M_{dyn}/M_{\odot}) - 11.0) $$ with
$\epsilon_y = 0.34$ and a scatter of $\pm 0.39$ dex; as with the other correlations, we have superimposed 
the solutions on the relevant data as shown in Figure 8.
We find, as in previous studies, the same near-linearity for \mbh versus $M_{dyn}$.
However, the residual dispersions and the required $\epsilon_y$ are slightly larger than for
\ngc - $M_{dyn}$.

\subsubsection[]{\ngc Versus \mbh}

Lastly, we plot \ngc vs \mbh and the inverse in Figures 9 and 10.  For the ``E+S'' solutions which we view as
the most reliable (excluding the S0 types whose apparent lack of correlation is not yet
understood), we obtain
$$ {\rm log} M_{\bullet} \, = \, (8.273 \pm 0.059) + (0.980 \pm 0.090) ({\rm log} N_{GC} - 2.7) $$ 
with $\epsilon_y = 0.30$ and a scatter of $\pm 0.34$ dex.
For the inverse,
$$ {\rm log} N_{GC} \, = \, (2.968 \pm 0.053) + (0.830 \pm 0.069) ({\rm log} M_{\bullet} - 8.5) $$ 
with $\epsilon_y = 0.27$ and a scatter of $\pm 0.30$ dex.

\subsection[]{Optimized Solutions}

\begin{table*}
\caption{Optimized Solutions from $\chi^2$ Minimization}
\label{Tab3}
\begin{tabular}{llccccc}
\hline
Solution For: & Sample & N & $\alpha$ & $\beta$ & $\langle x \rangle $ & RMS $\sigma_y$  \\
\hline
log $N_{GC}$ vs log $\sigma_e$ & E & 43 & $2.790 \pm 0.062$ & $4.600 \pm 0.088$ & 2.30 & 0.383  \\
log $N_{GC}$ vs log $M_{dyn}$ & E & 46  & $2.931 \pm 0.051$ & $0.940 \pm 0.072$ & 11.0 & 0.295  \\
log $N_{GC}$ vs log \mbh & E & 41  & $2.964 \pm 0.052$ & $0.840 \pm 0.070$ & 8.5 & 0.318   \\
log \mbh vs log $\sigma_e$ & E & 49 & $8.406 \pm 0.071$ & $5.580 \pm 0.422$ & 2.30 & 0.469  \\
log \mbh vs log $M_{dyn}$ & E & 44 & $8.541 \pm 0.061$ & $1.140 \pm 0.079$ & 11.0 & 0.323   \\
log \mbh vs log $N_{GC}$ & E   & 41 & $8.185 \pm 0.063$ & $1.190 \pm 0.027$ & 2.70 & 0.344   \\
\hline
\end{tabular}
\end{table*}

With solutions in hand for several pairs of quantities selected from
$M_{dyn}, N_{GC}$, \mbh, and $\sigma_e$, it should be possible in principle
to determine a value of $\epsilon$ (the cosmic scatter) for each one that
would yield self-consistent solutions ($\chi_{\nu}^2 \simeq 1$)
for every pair.  After experimentation with different $\epsilon$'s, we find 
that a self-consistent set is achieved near \\
$\epsilon$(log $M_{dyn}$) = 0.2 dex; \\
$\epsilon$(log \mbh) = 0.25 dex; \\
$\epsilon$(log $\sigma_e$) = 0.07 dex; \\
$\epsilon$(log $N_{GC}$)  = 0.2 dex. \\
We repeat that these are the additional scatters \emph{over and above} the measurement
uncertainties quoted in the literature that are needed to obtain correct 
$\chi^2-$minimization solutions.  With these four `optimal' $\epsilon$'s
the resulting correlations are as given in Table 3.

For each pair, we use the samples labelled ``All'' in Table 2.
In some cases the best-fit slopes change noticeably from the solutions given in Table 2
(where we had put all the additional scatter into $\epsilon_y$).  These differences 
arise directly from the changes in the relative weightings of the datapoints when
the assumed cosmic scatters are shared between the two quantities.

\subsection[]{MCMC Analysis}

To test for systematics in our fitting results, we perform a complementary analysis using a 
standard Monte-Carlo Markov Chain (MCMC) approach.  We implement the Metropolis-Hastings method 
\citep{hastings70} using a codebase which has been well tested against a diverse range of 
problems in several published works \citep[e.g.][]{Poole2013,Mutch2013}.  
Here, we present a number of solutions, each being a joint fit for the 
log-log slopes and intercepts of the \mdyn - \ngc, \mdyn - \sige, \mbh - \ngc, 
and \mbh - \sige relations.  We use Eqn. \ref{equation-chi2} to calculate 
model likelihoods assuming that uncertainties in the observed properties presented in Table 1 are normally 
distributed for \sige and \ngc and log-normally distributed for \mdyn and \mbh.  
In all cases, chains are calculated from $10^5$ burn-in iterations and $2{\times }10^6$ 
integration iterations.  Proposal selection is optimised with a rotated covariance matrix computed iteratively 
pre-burn-in from $2000$ proposals until all matrix elements converge to within $5\%$.  
Step sizes are selected to ensure a success rate of approximately 35\% throughout.  

All results presented in this section omit the Milky Way and NGC4486A,B for the same 
reasons given above.  Lastly, all fits presented in this section have $n_{DoF}$=36 degrees 
of freedom for which the $p$-value statistic rejects the null-hypothesis at a significance 
of $\alpha(p){\le}0.05$ for $\chi^2/n_{DoF}{\le}50.997{/}36{=}1.42$.  
We will use this throughout as our criterion for assessing the goodness of fit for
the MCMC solutions.

\begin{figure*}
\begin{minipage}{190mm}
\begin{center}
 \includegraphics[width=160mm]{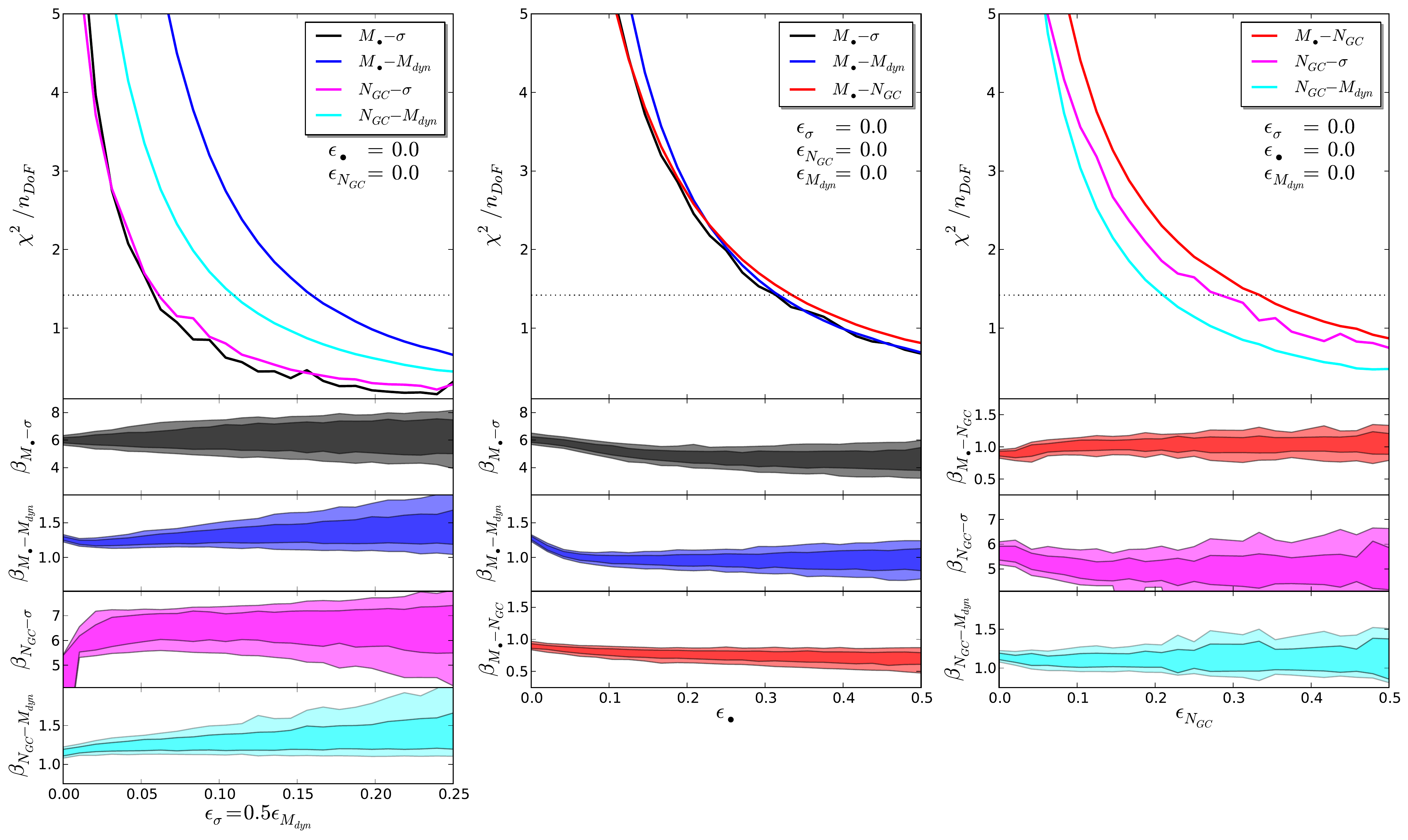}
 \\
 \includegraphics[width=160mm]{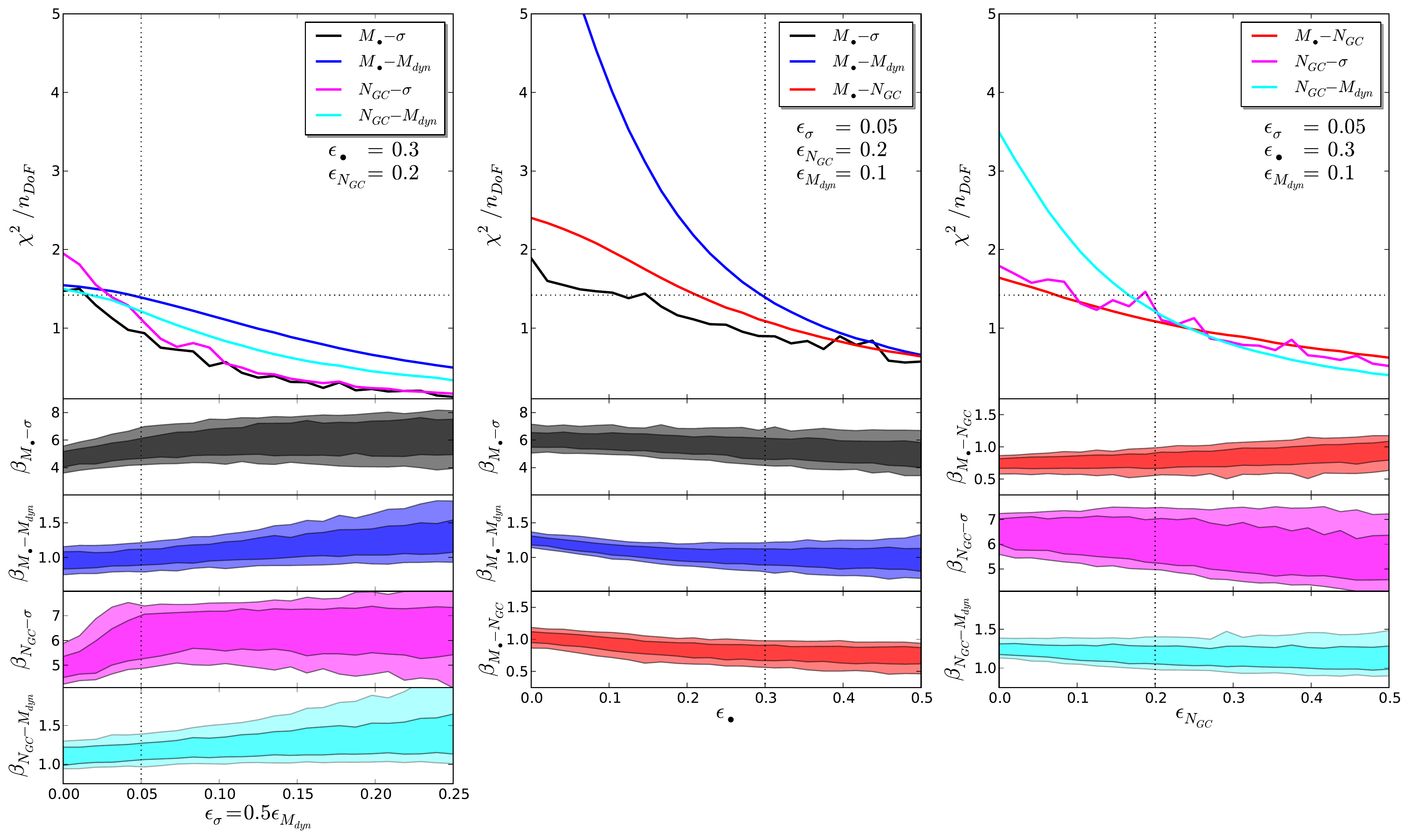}
 \caption{The effects of accounting for increased measurement uncertainties 
or intrinsic scatter on the goodness of fit and best-fit slopes of our 5 scaling 
relations of interest.  Goodness of fit is expressed in the top frame of each 
column of plots (expressed in terms of reduced-$\chi^2$ for 36 degrees of freedom) 
with a horizontal dotted line depicting our good-fit criterion ($\chi^2/n_{DoF}{\le}1.42$).  
The 68\% and 95\% confidence intervals for our best fit slopes are shown with dark and light 
coloured shaded regions respectively.  The top set of plots shows the effect of each scatter 
variable ($\epsilon_\sigma$,$\epsilon_{M_{dyn}}$,$\epsilon_{M_\bullet}$ or $\epsilon_{N_{GC}}$) 
in isolation while the bottom set shows the same but for a self-consistent set of values 
permitting a good fit to every relation (indicated with vertical dotted lines).  Since the effects of $\epsilon_\sigma$ on relations 
involving $M_{dyn}$ is mathematically equivalent to the effects of 
$\epsilon_{M_{dyn}}$ through the relation $\epsilon_{M_{dyn}}{=}2\epsilon_\sigma$, 
we present both on the same axis to compress our results.}
 \label{fig-chi2}
\end{center}
\end{minipage}
\end{figure*}

To explore the idea that some or all of the uncertainties presented in Table 1 have 
been systematically underestimated or that the underlying relations host some degree of intrinsic 
scatter, we have applied this fitting procedure for various values of $\epsilon_x$ 
and $\epsilon_y$ for each relation.  In the top panel of Figure \ref{fig-chi2} 
we present the influence of independently changing the value of $\epsilon$ for each observable 
of interest ($\epsilon_\sigma$, $\epsilon_{M_{dyn}}$, $\epsilon_\bullet$ and $\epsilon_{N_{GC}}$) on 
the resulting reduced-$\chi^2$ and on the inferred slopes $\beta$ for each relation.  
Since the effect of $\epsilon_\sigma$ on relations involving $M_{dyn}$ is mathematically 
equivalent to the effects of $\epsilon_{M_{dyn}}$ through the relation 
$\epsilon_{M_{dyn}}{=}2\epsilon_\sigma + \epsilon_{R_e}$, 
we present both on the same axis to compress our results.  In this way, 
the effects of $\epsilon_{M_{dyn}}$ can be interpreted in terms of underestimated uncertainties 
in $\sigma$ (which carries through to estimates of $M_{dyn}$ through Eqn. \ref{eqn-M_dyn}) or 
separately as intrinsic scatter or underestimated uncertainties in the measurement of $M_{dyn}$.

The results from these calculations agree generally well with those presented 
in the previous section: for every relation we find that fits of acceptable quality can only 
be achieved when $\epsilon_x$ and/or $\epsilon_y$ are significantly larger than zero. 
In addition the slopes $\beta$ change with $\epsilon$, though how much a change
is generated differs from one pair of parameters to another  and sometimes moves 
in opposite directions for $\epsilon_x$ and $\epsilon_y$.

For relations whose slope changes significantly with increasing values 
of $\epsilon_x$ or $\epsilon_y$, most of this change occurs at small initial displacements
away from $\epsilon = 0$, 
converging to constant values by the time the $\epsilon$'s have increased sufficiently to permit a good fit.  
This trend is consistent with a scenario where a small number of precisely measured data points are dominating the fit.
Our results suggest that $\epsilon_\sigma$ is small ($\le$5\%) while the other $\epsilon$'s 
need to be of the order of $10{-}30$\%.
Moreover, covariance between our scaling relations demands that the $\epsilon$-values required 
for good fits in Figure \ref{fig-chi2} represent upper limits to what the lower limits 
of these values must be  
(since fits with $\chi^2/n_{DoF}{<}1.42$ are considered good as well).

Taking the values of $\epsilon_\sigma{\sim}0.05$, $\epsilon_\bullet{\sim}0.3$, 
$\epsilon_{N_{GC}}{\sim}0.2$ suggested by the top set of plots in this figure and iterating slightly 
for $\epsilon_{M_{dyn}}$, we have manually determined a self-consistent set of minimum 
values $(\epsilon_\sigma,\epsilon_\bullet,\epsilon_{N_{GC}},
\epsilon_{M_{dyn}}){=}(0.05,0.30,0.20,0.1)$.  These particular values 
yield a good fit for each scaling relation of interest.
In addition, these four values are encouragingly consistent with the optimal set of values we
obtained in section 4.2 above through our $\chi^2$ minimization solutions.
This result is shown in the bottom set of plots in Figure \ref{fig-chi2} where we see 
that every relation converges to a good fit at approximately this set of values.  
The previous trends of $\beta_{x-y}$ with $\epsilon_x$ and $\epsilon_y$ are still seen, but the 
amplitude of the dependence is generally reduced and the trends with $\epsilon$ less severe.
This dependence of $\beta_{x-y}$ on $\epsilon_x$ and $\epsilon_y$ reliably moves in 
opposite directions for $\epsilon_x$ and $\epsilon_y$ now.
And, furthermore, results for the slopes are insensitive to modest changes 
in the set of $\epsilon$ values chosen.
Finally and importantly, increases in the $\epsilon$ values leading to 
reduced-$\chi^2{\ll}1.42$ do not lead to significant changes of slopes.

\begin{figure*}
\begin{minipage}{190mm}
\begin{center}
 \includegraphics[width=180mm]{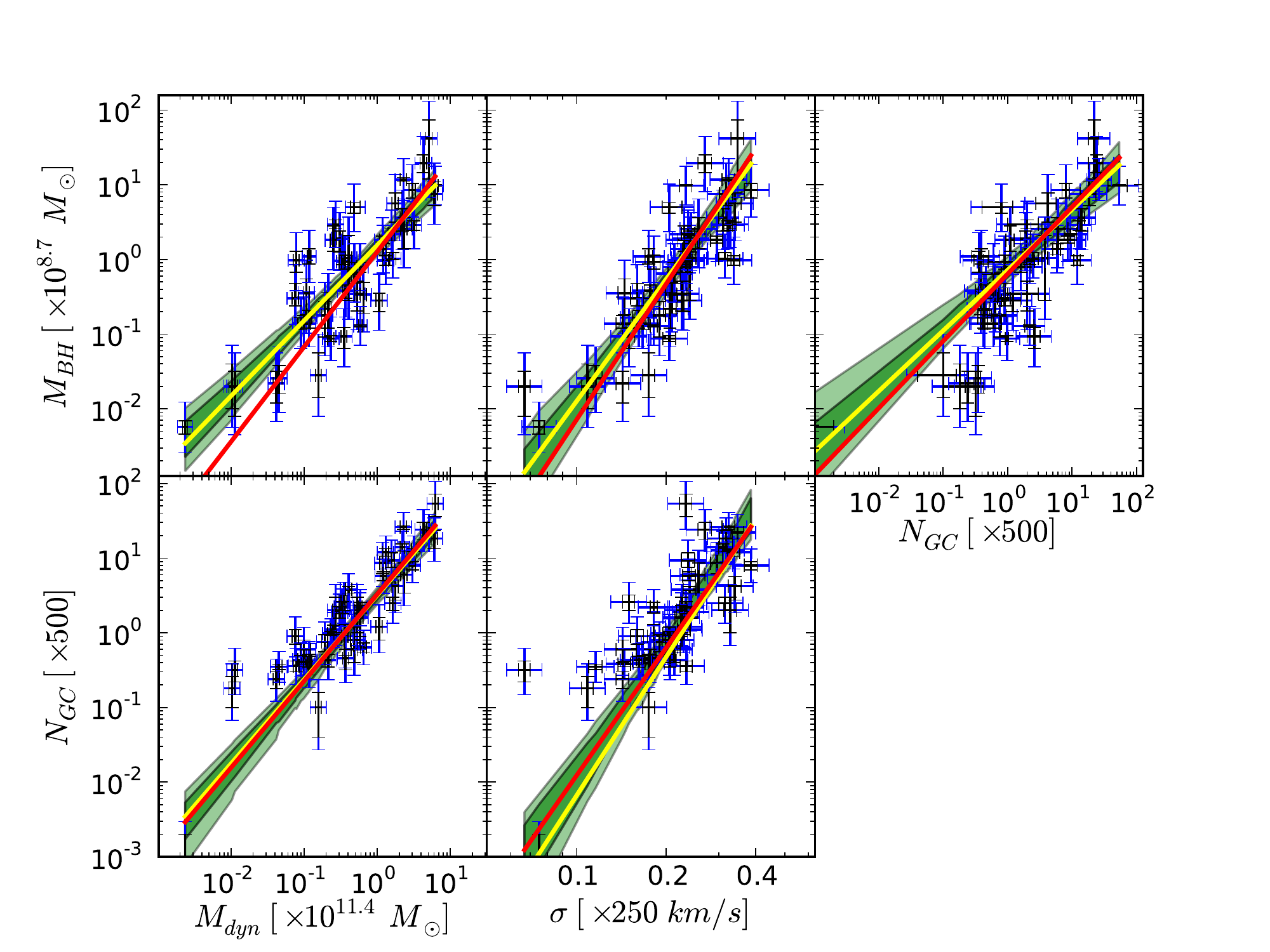}
 \caption{MCMC fits to the 2-parameter scaling relations.
The upper panels shows \mbh vs. \mdyn, $\sigma_e$, and \ngc, while
the lower panels show \ngc vs. \mdyn and $sigma_e$.  In each panel the
{\sl yellow line} marks the `optimal' solution where both coordinates
are allowed to have additional cosmic scatter and the shaded green areas
mark the 68\% and 95\% confidence limits on the solution.  The
{\sl red line} denotes the best-fit solution where only the y-axis is
allowed to have additional scatter.}\label{fig-fits_plot}
\end{center}
\end{minipage}
\end{figure*}

The fits we derive using this set of optimal values for $\epsilon$ are listed in 
Table \ref{table-MCMC_fits} and are depicted in Figure \ref{fig-fits_plot}.
In each panel of tthe Figure, the heavy yellow line shows the optimal solution,
with 68\% and 95\% confidence intervals shown as dark and light green shaded regions. 
Errorbars in black show the individual measurement uncertainties in the literature
(no $\epsilon$'s added), while the effects of additional 
scatter are depicted in the blue errorbars.  The heavy red line shows the best fit derived 
from $\epsilon_x = 0$ for comparison, primarily to show how much the sharing of cosmic
scatter among the parameters generates slope changes.

Notably, the addition of scatter (i.e. adopting the optimal solutions) has the effect of flattening all scaling 
relations involving the black-hole mass $M_\bullet$.  This flattening is marginally significant at 
a ${\sim}68$\% level in the case of the $M_\bullet{-}\sigma$ and $M_\bullet{-}N_{GC}$ 
relations, but is highly significant at ${\gg}95$\% confidence for the $M_\bullet{-}M_{dyn}$ relation.
By contrast, the addition of scatter has no significant effect on the resulting slopes for relations involving $N_{GC}$.

\begin{table*}
\caption{Correlation solutions from our MCMC analysis.  Unbracketed quantities are $68$\% confidence 
results and bracketed quantities are $95$\% confidence results.}
\label{table-MCMC_fits}
\begin{tabular}{lclrlrl}
\hline
\multicolumn{3}{c}{Scaling Relation} &
\multicolumn{2}{c}{$\alpha$} &
\multicolumn{2}{c}{$\beta$} \\
\hline
 log $N_{GC}$ &vs& log $\sigma_e$ & $ 4.29^{+0.49}_{-0.70}$ & ($^{+0.70}_{-0.94}$) & $6.11^{+0.90}_{-0.86}$ & ($^{+1.32}_{-1.24}$) \\
 log $N_{GC}$ &vs& log $M_{dyn}$  & $ 0.53^{+0.05}_{-0.09}$ & ($^{+0.11}_{-0.14}$) & $1.13^{+0.16}_{-0.07}$ & ($^{+0.28}_{-0.14}$) \\
 log $N_{GC}$ &vs& log \mbh       & $-0.14^{+0.11}_{-0.12}$ & ($^{+0.20}_{-0.20}$) & $0.81^{+0.10}_{-0.14}$ & ($^{+0.17}_{-0.25}$) \\
 log \mbh &vs& log $\sigma_e$     & $ 3.58^{+0.40}_{-0.60}$ & ($^{+0.87}_{-0.88}$) & $5.36^{+0.76}_{-0.70}$ & ($^{+1.56}_{-1.16}$) \\
 log \mbh &vs& log $M_{dyn}$      & $ 0.20^{+0.08}_{-0.11}$ & ($^{+0.15}_{-0.17}$) & $1.01^{+0.11}_{-0.12}$ & ($^{+0.20}_{-0.22}$) \\
\hline
\end{tabular}
\end{table*}

Table 3 (classic $\chi^2$ minimization) should be compared with Table 4 (MCMC fits).
The differences in the $\alpha-$ values are due simply to adoption of different
zeropoint scalings.  The more important comparison between the deduced slopes $\beta$
reveals mutual agreement to within the combined uncertainties for most of the 
parameter pairs.  Given the level of total scatter in all the relations, and
the limitations in the still-small sample of galaxies, we suggest that the MCMC
uncertainties on the slopes are likely to be the more realistic of the two methods.
The largest discrepancy is nominally for \ngc - \sigea (slope 4.86 from Table 3, 6.11
from Table 4), but since the combined uncertainty is also large ($\sim \pm 1.2$) the
difference does not stand out especially 
when we also consider the presence of some individual outliers and the fact that
the relation is intrinsically nonlinear (see again HHA13).
The correlation for \mbh - \ngc yielded a noticeably flatter slope from
MCMC ($\simeq 0.81 \pm 0.2$) than from $\chi^2-$minimization ($\simeq 1.19 \pm 0.03$),
apparently because MCMC was less influenced by the dwarf-galaxy points appearing
at the lower left of the graph.  

The message we take from these comparisons is
that there is no evidence at the present time to reject the view that
\mbh and \ngc are rather closely directly proportional to one another.
Both of these, in turn, scale almost exactly as \mdyna.

\section{Summary and Conclusions}

We have taken advantage of recent compilations of global parameters for
galaxies in which the mass of the central black hole  
has been determined, and have combined these with a recent compilation of all galaxies with known globular
cluster populations.  We now find
a sample of 49 galaxies which can be used to explore the \mbh - \ngc correlation
that has attracted recent interest.
For this set of galaxies we also have values of \re and \sige which can be used to examine 
correlations of \mbh and \ngc with \sige and \mdyna.  

Examining the raw data in detail,
we find that particularly for \re and \sige, 
wide ranges of values for a given galaxy are often quoted in the literature that make it difficult to
select best measurements.  For these parameters we chose to use the values from the \mbh determinations as an 
attempt at consistency.  All parameters were renormalized (where appropriate) to 
the \citet{ton01} SBF galaxy distance catalogue as it contained virtually all of the galaxies in our sample and 
introduced no systematic biases within the quoted errors.

With the data available, we have solved for correlations among the parameters \mbha,
\ngca, \sigea, and \mdyn using both $\chi^2$ minimization and Monte Carlo Markov Chain
techniques.

\begin{itemize}
\item All the 2-parameter correlations show total scatters around the best-fit relations
of $\pm0.3 - 0.4$ in the log.  Both \ngc and \mbh follow tight and nearly linear
relations particularly with \mdyn (bulge mass). Among all parameter pairs, the single relation with the lowest absolute scatter is
\ngc versus \mdyna; these two quantities scale accurately in direct proportion,
a result that we have also found from much larger galaxy samples (HHA13).
\item One of our strongest findings is that in both fitting methods we must introduce additional variance 
$\epsilon$ over and above the formal measurement uncertainties in order to obtain statistically valid solutions.
To varying degrees these additional variances are needed in every one of the 4 parameters we use.
The clear implication is that either the galaxy
parameters have significant `cosmic scatter', or (perhaps less likely) that the observational measurement
uncertainties have been significantly underestimated.
By using all of the 2-parameter correlations that are obtainable from the data, we find a self-consistent 
set of {\it extra variances} (in log scale) given by 
$(\epsilon_\sigma,\epsilon_\bullet,\epsilon_{N_{GC}},\epsilon_{M_{dyn}}){=}(0.05,0.30,0.20,0.1)$.  
The \mbh values show the largest extra variance.
As mentioned above, the need for additional variation in the correlation solutions is not new (e.g. \citet{mf01},
 \citet{t1}, \citet{novak06} among others).  What is new here is that the additional variance is required
in all four parameters.  For comparison we determined average measurement uncertainties for
\sigea, \mbha, and \ngc based on the data in Table 1.  In this exercise we excluded the galaxies
M32, NGC 4486A, and NGC 4486B (which were commonly left out of the correlation solutions) as well 
as NGC4374 and A2052 for which the quoted values of \mbh were upper limits only.  We used 
$\epsilon_{M{dyn}}$ = 0.1as discussed in section 3.6.  The variances for each parameter, based only on the quoted observational errors, are then
$(\epsilon_\sigma,\epsilon_\bullet,\epsilon_{N_{GC}},\epsilon_{M_{dyn}}){=}(0.02,0.10,0.19,0.1)$ in the log.  
In all cases (except for $\epsilon_{M_{dyn}}$), the {\it extra} variance required is more than twice
the nominal observed values.  
\item The correlations involving \mbh have slopes that are more sensitive to the addition of
$\epsilon$'s than those involving \ngca; the additional variance increases the relative importance
of the smaller galaxies to the fit and ends up flattening all the slopes.
\item The correlation of \ngc against \sige is poorly constrained and (based on solutions from
larger samples) more strongly nonlinear than the other parameter pairs.
\end{itemize}

At this stage of development, we suggest that the best observational route to exploring
the \ngc - \mbh correlation further will be to reduce sample bias by adding more
galaxies at lower luminosity, but especially by adding more
S and S0 galaxies; at present, the E galaxies dominate the still-limited sample
of galaxies in which both \ngc and \mbh are accurately measured.

The point of exploring these and other correlations among galaxy parameters
is to gain an understanding of galaxy formation and evolution.  All these parameters
arise from the original mass concentration and gravitational potential well
from which each galaxy formed.  But each parameter examined here (as well as those 
such as galaxy luminosity, mass and halo mass among others) has evolved with time and 
in different ways, changing the interrelations as well.  For instance, \mbh is expected 
to increase as gas and stars surrounding the galaxy centre are accreted and, in many 
cases this process is ongoing today.  Contrarily \ngc will decrease with time as 
clusters are disrupted by both internal and external effects; but, in contrast to \mbha, 
most of the change in \ngc is expected to have happened within the first $\sim 10^9$ years
after formation.  \ngc evolution is more smoothly predictable and due to the overall tidal
field while \mbh evolution is more stochastic and can differ even between similar galaxies due 
to local differences in gas-rich merger histories. Thus, it is not surprising that \mbh 
relations have greater scatter than those involving \ngca.
The physical scales at work are very different as well, from $\sim 1$pc for
the central black hole to $\sim 10^5$pc for the globular cluster system.  Since it is
likely that most or all galaxy properties have changed over the past $\sim 10^{10}$ years, 
then the nature of the various correlations today (their slope, linearity or curvature, 
and dispersion) will reflect {\it both} initial conditions and evolution to some
degree.  

Lastly, our results suggest that the dispersions observed in the various
correlations are significantly affected
by the individual evolutionary histories of the galaxies
 and cannot be explained by saying that the quoted uncertainties for 
the many galaxy properties are too small.
We therefore favor the interpretation that the extra variances, $\epsilon$, introduced by both 
the reduced $\chi^2$ and MCMC analyses are more likely to be reasonable estimates of 
the cosmic scatter.

\section*{Acknowledgments}

This publication makes use of data products from the Two Micron All-Sky Survey which is
a joint project of the University of Massachusetts and the Infrared Processing and Analysis Center/California
Institute of Technology, funded by the National Aeronautics and Space Administration
and the National Science Foundation.

We thank John Kormendy, Alister Graham, James Taylor, and Brian McNamara for helpful discussions.  
Much of this work was begun during an extended visit to Mount Stromlo Observatory
(Australian National University).  GP acknowledges support from S. Wyithe's ARC Laureate 
Grant (FL110100072).
WEH also acknowledges financial 
support through a grant from the Natural Sciences and Engineering Research 
Council of Canada.  Lastly, we thank the referee for comments which helped 
improve the clarity of our discussion.

\label{lastpage}

\end{document}